# The complexity of conservative valued CSPs[*]

Vladimir Kolmogorov[†]

Institute of Science and Technology (IST), Austria

vnk@ist.ac.at

Stanislav Živný[‡]

University of Oxford, UK

standa.zivny@cs.ox.ac.uk

## Abstract

We study the complexity of valued constraint satisfaction problems (VCSP). A problem from VCSP is characterised by a *constraint language*, a fixed set of cost functions over a finite domain. An instance of the problem is specified by a sum of cost functions from the language and the goal is to minimise the sum. Under the unique games conjecture, the approximability of finite-valued VCSPs is well-understood, see Raghavendra [FOCS'08]. However, there is no characterisation of finite-valued VCSPs, let alone general-valued VCSPs, that can be solved exactly in polynomial time, thus giving insights from a combinatorial optimisation perspective.

We consider the case of languages containing all possible unary cost functions. In the case of languages consisting of only $\{0, \infty\}$-valued cost functions (i.e. relations), such languages have been called *conservative* and studied by Bulatov [LICS'03] and recently by Barto [LICS'11]. Since we study valued languages, we call a language *conservative* if it contains all finite-valued unary cost functions. The computational complexity of conservative valued languages has been studied by Cohen *et al.* [AIJ'06] for languages over Boolean domains, by Deineko *et al.* [JACM'08] for $\{0, 1\}$-valued languages (a.k.a Max-CSP), and by Takhanov [STACS'10] for $\{0, \infty\}$-valued languages containing all finite-valued unary cost functions (a.k.a. Min-Cost-Hom).

We prove a Schaefer-like dichotomy theorem for conservative valued languages: if all cost functions in the language satisfy a certain condition (specified by a complementary combination of *STP and MJN multimorphisms*), then any instance can be solved in polynomial time (via a new algorithm developed in this paper), otherwise the language is NP-hard. This is the *first* complete complexity classification of *general-valued constraint languages* over non-Boolean domains. It is a common phenomenon that complexity classifications of problems over non-Boolean domains is significantly harder than the Boolean case. The polynomial-time algorithm we present for the tractable cases is a generalisation of the submodular minimisation problem and a result of Cohen *et al.* [TCS'08].

Our results generalise previous results by Takhanov [STACS'10] and (a subset of results) by Cohen *et al.* [AIJ'06] and Deineko *et al.* [JACM'08]. Moreover, our results do not rely on any computer-assisted search as in Deineko *et al.* [JACM'08], and provide a powerful tool for proving hardness of finite-valued and general-valued languages.

## 1 Introduction

The constraint satisfaction problem is a central generic problem in computer science. It provides a common framework for many theoretical problems as well as for many real-life applications, see [29] for a nice survey. An instance of the *constraint satisfaction problem* (CSP) consists of a collection of variables which must be assigned values subject to specified constraints. CSP is equivalent to the problem of evaluating conjunctive queries on databases [36], and to the homomorphism problem for relational structures [24].

---

[*]An extended abstract of this work will appear in the *Proceedings of the 23rd Annual ACM-SIAM Symposium on Discrete Algorithms (SODA)*, 2012.

[†]Vladimir Kolmogorov was supported by the Royal Academy of Engineering/EPSRC, UK.

[‡]Stanislav Živný is supported by a Junior Research Fellowship at University College, Oxford. Part of this work was done while the second author was visiting Microsoft Research Cambridge.

An important line of research on the CSP is to identify all tractable cases; that is, cases that are recognisable and solvable in polynomial time. Most of this work has been focused on one of the two general approaches: either identifying structural properties of the way constraints interact which ensure tractability no matter what forms of constraints are imposed [22], or else identifying forms of constraints which are sufficiently restrictive to ensure tractability no matter how they are combined [11, 24].

The first approach has been used to characterise all tractable cases of bounded-arity CSPs: the *only* class of structures which ensures tractability (subject to a certain complexity theory assumption, namely FPT $\neq$ W[1]) are structures of bounded tree-width modulo homomorphic equivalence [20, 26, 27, 39]; and recently also for unbounded-arity CSPs [40]. The second approach has led to identifying certain algebraic properties known as polymorphisms [32] which are necessary for a set of constraint types to ensure tractability. A set of constraint types which ensures tractability is called a *tractable constraint language*.

Schaefer in his seminal work [44] gave a complete complexity classification of Boolean constraint languages. The algebraic approach based on polymorphisms [33] has been so far the most successful tool in generalising Schaefer's result to languages over a 3-element domain [10], languages with all unary relations [12, 4], languages comprised of a single binary relation without sources and sinks [3] (see also [6]), and languages comprised of a single binary relation that is a special triad [2]. The algebraic approach has also been essential in characterising the power of local consistency [5] and the "few subpowers property" [7, 30], the two main tools known for solving tractable CSPs. A major open question in this line of research is the *Dichotomy Conjecture* of Feder and Vardi, which states that every constraint language is either tractable or NP-hard [24]. We remark that there are other approaches to the dichotomy conjecture; see, for instance, [29] for a nice survey of Hell and Nešetřil, and [37] for a connection between the Dichotomy Conjecture and probabilistically checkable proofs.

Since in practice many constraint satisfaction problems are over-constrained, and hence have no solution, or are under-constrained, and hence have many solutions, *soft* constraint satisfaction problems have been studied [21]. In an instance of the soft CSP, every constraint is associated with a cost function (rather than a relation as in the CSP) which represents preferences among different partial assignments, and the goal is to find the best assignment. Several very general soft CSP frameworks have been proposed in the literature [45, 9]. In this paper we focus on one of the very general frameworks, the *valued* constraint satisfaction problem (VCSP) [45]. Throughout the paper, we use the term *constraint language* (or just *language*) for a set of cost functions over a finite domain. If all cost functions from a given language $\Gamma$ are $\{0, \infty\}$-valued (i.e. relations), we call $\Gamma$ a *crisp* language. (If necessary, to stress the fact that $\Gamma$ is a language, but not a crisp language, we call $\Gamma$ a *general-valued* language.)

Similarly to the CSP, an important line of research on the VCSP is to identify tractable cases which are recognisable in polynomial time. Is is well known that structural reasons for tractability generalise to the VCSP [8]. In the case of language restrictions, only a few conditions are known to guarantee tractability of a given language [15, 14].

**Related work** The problem of characterising the complexity of different languages has received significant attention in the literature. For some classes researchers have established a Schaefer-like dichotomy theorem of the following form: if language $\Gamma$ admits certain *polymorphisms* or *multimorphisms* then it is tractable, otherwise it is NP-hard. Some of these classes are as follows: Boolean languages, i.e. languages with a 2-element domain (Cohen *et al.* [15]); crisp languages including all unary relations (Bulatov [12] and recently Barto [4]); crisp languages with a 3-element domain (Bulatov [10]); $\{0, 1\}$-valued languages including all unary cost functions (Deineko *et al.* [23]); crisp languages including additionally all finite-valued unary cost functions (Takhanov [46]); crisp languages including additionally a certain subset of finite-valued unary cost functions (Takhanov [47]).

Our proof exploits the results of Takhanov [46], who showed the existence of a majority polymorphism as a necessary condition for tractability of crisp languages including additionally all finite-valued unary cost functions. Other related work includes the work of Creignou *et al.* who studied various generalisations of the CSP to optimisation problems over Boolean domains [18], see also [19, 35]. Raghavendra [42] and Raghavendra and Steurer [43] have shown how to optimally approximate any finite-valued VCSP.

**Contributions** This paper focuses on valued languages containing all finite-valued unary cost functions; we call such languages *conservative*. Our main result is a dichotomy theorem for all conservative languages: if a conservative language $\Gamma$ admits a complementary combination of *STP (symmetric tournament pair) and MJN (majority-majority-minority) multimorphisms*, then it is tractable, otherwise $\Gamma$ is NP-hard. This is the first complete complexity classification of general-valued languages over non-Boolean domains, generalising previously obtained results in [15, 23, 46] as follows:

- Cohen *et al.* proved a dichotomy for arbitrary Boolean languages ($|D| = 2$). We generalise it to arbitrary domains ($|D| \geq 2$), although only for conservative languages.

- Deineko *et al.* [23] and Takhanov [46] proved a dichotomy for the following languages, respectively:
  - $\{0, 1\}$-valued languages containing additionally all unary cost functions;
  - $\{0, \infty\}$-valued languages containing additionally all unary cost functions.

  In both of these case the languages are conservative, so these classifications are special cases of our result. Note, however, that Deineko *et al.* additionally give a dichotomy with respect to approximability (PO vs. APX-hard), even when the number of occurrences of variables in instances is bounded; this part of [23] does not follow from our classification.

Moreover, our results provide a new powerful tool and do not rely on a computer-assisted search as in [23]. Building on techniques from this paper, Jonsson *et al.* [34] have recently shown that the same approach can be also used for certain non-conservative languages.

Since the complexity of Boolean conservative languages is known, we start, similarly to Bulatov and Takhanov [12, 46], by exploring the interactions between different 2-element subdomains. Given a conservative language $\Gamma$, we will investigate properties of a certain graph $G_\Gamma$ associated with the language and cost functions expressible over $\Gamma$. We link the complexity of $\Gamma$ to certain properties of the graph $G_\Gamma$.

First, we show that if $G_\Gamma$ does not satisfy certain properties, then $\Gamma$ is intractable. Second, using $G_\Gamma$, we construct a (partial) *STP multimorphism* and a (partial) *MJN multimorphism*. Finally, we show that any language which admits a complementary combination of *STP and MJN multimorphisms* is tractable, thus generalising a tractable class of Cohen *et al.* [14], which in turn is a generalisation of the submodular minimisation problem. Thus we obtain a dichotomy theorem. The general-valued case is much more involved than the finite-valued case, and requires different techniques compared to previous results.

Given a finite language $\Gamma$, the graph $G_\Gamma$ is finite as well, but depends on the expressive power of $\Gamma$ (see Section 2 for precise definitions), which is infinite. In order to test whether $\Gamma$ is tractable, we do not need to construct the graph $G_\Gamma$ as it follows from our result that we just need to test for the existence of a complementary combination of two multimorphisms, which can be established in polynomial time.

Our results are formulated using the terminology of valued constraint satisfaction problems, but they apply to various other optimisation frameworks that are equivalent to valued constraint satisfaction problems such as Gibbs energy minimisation, Markov Random Fields, Min-Sum problems, and other models [38, 49].

**Organisation of the paper** The rest of the paper is organised as follows. In Section 2, we define valued constraint satisfaction problems (VCSPs), conservative languages, multimorphisms and other necessary definitions needed throughout the paper. We state our results in Section 3, and then give their proofs in Sections 4-7.

## 2 Background and notation

We denote by $\mathbb{Q}_+$ the set of all non-negative rational numbers. We define $\overline{\mathbb{Q}}_+ = \mathbb{Q}_+ \cup \{\infty\}$ with the standard addition operation extended so that for all $a \in \mathbb{Q}_+$, $a + \infty = \infty$. Members of $\overline{\mathbb{Q}}_+$ are called *costs*. Throughout the paper, we denote by $D$ any fixed finite set, called a *domain*. Elements of $D$ are called *domain values* or *labels*.

A function $f$ from $D^m$ to $\overline{\mathbb{Q}}_+$ will be called a *cost function* on $D$ of *arity* $m$. If the range of $f$ lies entirely within $\mathbb{Q}_+$, then $f$ is called a *finite-valued* cost function. If the range of $f$ is $\{0, \infty\}$, then $f$ is called a *crisp* cost function. If the range of a cost function $f$ includes non-zero finite costs and infinity, we emphasise this fact by calling $f$ a *general-valued* cost function. Let $f : D^m \to \overline{\mathbb{Q}}_+$ be an $m$-ary cost function $f$. We denote $\text{dom} f = \{\boldsymbol{x} \in D^m \mid f(\boldsymbol{x}) < \infty\}$ to be the effective domain of $f$. The argument of $f$ is called an *assignment* or a *labelling*. Functions $f$ of arity $m = 2$ are called *binary*.

A *language* is a set of cost functions with the same domain $D$. Language $\Gamma$ is called finite-valued (crisp, general-valued respectively) if all cost functions in $\Gamma$ are finite-valued (crisp, general-valued respectively). A language $\Gamma$ is *Boolean* if $|D| = 2$.

**Definition 1.** *An instance $\mathcal{I}$ of the* valued constraint satisfaction problem *(VCSP) is a function $D^V \to \overline{\mathbb{Q}}_+$ given by*

$$\text{Cost}_\mathcal{I}(\boldsymbol{x}) = \sum_{t \in T} f_t \left( x_{i(t,1)}, \ldots, x_{i(t,m_t)} \right)$$

*It is specified by a finite set of nodes $V$, finite set of terms (also known as constraints) $T$, cost functions $f_t : D^{m_t} \to \overline{\mathbb{Q}}_+$ or arity $m_t$ and indices $i(t, k) \in V$ for $t \in T$, $k = 1, \ldots, m_t$. A solution to $\mathcal{I}$ is an assignment $\boldsymbol{x} \in D^V$ with the minimum cost.*

We denote by $\text{VCSP}(\Gamma)$ the class of all VCSP instances whose terms $f_t$ belong to $\Gamma$. A finite language $\Gamma$ is called *tractable* if $\text{VCSP}(\Gamma)$ can be solved in polynomial time, and *intractable* if $\text{VCSP}(\Gamma)$ is NP-hard. An infinite language $\Gamma$ is tractable if every finite subset $\Gamma' \subseteq \Gamma$ is tractable, and intractable if there is a finite subset $\Gamma' \subseteq \Gamma$ that is intractable.

The idea behind conservative languages is to contain all possible unary cost functions: Bulatov has called a crisp language $\Gamma$ conservative if $\Gamma$ contains all unary relations [12]. We are interested in valued languages containing all possible unary cost functions and hence define conservative languages as follows:

**Definition 2.** *Language $\Gamma$ is called* conservative *if $\Gamma$ contains all $\{0, 1\}$-valued unary cost functions $u : D \to \{0, 1\}$.*

Such languages have been studied by Deineko *et al.* [23] and Takhanov [46]. Note, we could have defined $\Gamma$ to be conservative if it contains all possible general-valued unary cost functions $u : D \to \overline{\mathbb{Q}}_+$. However, the weaker definition 2 will be sufficient for our purposes: it is shown in Section 4 that adding all possible unary cost functions $u : D \to \overline{\mathbb{Q}}_+$ to a conservative language $\Gamma$ does not change the complexity of $\Gamma$.

We now define polymorphisms, which have played a crucial role in the complexity analysis of crisp languages [33, 11].

**Definition 3.** *A mapping $F : D^k \to D$, $k \geq 1$ is called a* polymorphism *of a cost function $f : D^m \to \overline{\mathbb{Q}}_+$ if*

$$F(\boldsymbol{x}_1, \ldots, \boldsymbol{x}_k) \in \operatorname{dom} f \qquad \forall \boldsymbol{x}_1, \ldots, \boldsymbol{x}_k \in \operatorname{dom} f$$

*where $F$ is applied component-wise. $F$ is a polymorphism of a language $\Gamma$ if $F$ is a polymorphism of every cost function in $\Gamma$.*

Multimorphisms [15] are generalisations of polymorphisms. To make the paper easier to read, we only define binary and ternary multimorphisms as we will not need multimorphisms of higher arities.

**Definition 4.** *Let $\langle \sqcap, \sqcup \rangle$ be a pair of operations, where $\sqcap, \sqcup : D \times D \to D$, and let $\langle F_1, F_2, F_3 \rangle$ be a triple of operations, where $F_i : D \times D \times D \to D$, $1 \leq i \leq 3$.*

- *Pair $\langle \sqcap, \sqcup \rangle$ is called a (binary)* multimorphism *of cost function $f : D^m \to \overline{\mathbb{Q}}_+$ if*

$$f(\boldsymbol{x} \sqcap \boldsymbol{y}) + f(\boldsymbol{x} \sqcup \boldsymbol{y}) \leq f(\boldsymbol{x}) + f(\boldsymbol{y}) \qquad \forall \boldsymbol{x}, \boldsymbol{y} \in \operatorname{dom} f \qquad (1)$$

  *where operations $\sqcap, \sqcup$ are applied component-wise. $\langle \sqcap, \sqcup \rangle$ is a multimorphism of language $\Gamma$ if $\langle \sqcap, \sqcup \rangle$ is a multimorphism of every $f$ from $\Gamma$.*

- *Triple $\langle F_1, F_2, F_3 \rangle$ is called a (ternary)* multimorphism *of cost function $f : D^m \to \overline{\mathbb{Q}}_+$ if*

$$f(F_1(\boldsymbol{x}, \boldsymbol{y}, \boldsymbol{z})) + f(F_2(\boldsymbol{x}, \boldsymbol{y}, \boldsymbol{z})) + f(F_3(\boldsymbol{x}, \boldsymbol{y}, \boldsymbol{z})) \leq f(\boldsymbol{x}) + f(\boldsymbol{y}) + f(\boldsymbol{z}) \qquad \forall \boldsymbol{x}, \boldsymbol{y}, \boldsymbol{z} \in \operatorname{dom} f \quad (2)$$

  *where operations $F_1, F_2, F_3$ are applied component-wise. $\langle F_1, F_2, F_3 \rangle$ is a multimorphism of language $\Gamma$ if $\langle F_1, F_2, F_3 \rangle$ is a multimorphism of every $f$ from $\Gamma$.*

- *Operation $F : D^k \to D$ is called* conservative *if $F(x_1, \ldots, k_k) \in \{x_1, \ldots, x_k\}$ for all $x_1, \ldots, x_k \in D$.*

- *Pair $\langle \sqcap, \sqcup \rangle$ is called* conservative *if $\{\!\{a \sqcap b, a \sqcup b\}\!\} = \{\!\{a, b\}\!\}$ for all $a, b \in D$, where $\{\!\{\ldots\}\!\}$ denotes a* multiset*, i.e. in the case of repetitions elements' multiplicities are taken into account. Similarly, triple $\langle F_1, F_2, F_3 \rangle$ is called* conservative *if $\{\!\{F_1(a, b, c), F_2(a, b, c), F_3(a, b, c)\}\!\} = \{\!\{a, b, c\}\!\}$ for all $a, b, c \in D$. In other words, applying $\langle F_1, F_2, F_3 \rangle$ to $(a, b, c)$ should give a permutation of $(a, b, c)$.*

- *Pair $\langle \sqcap, \sqcup \rangle$ is called a* symmetric tournament pair (STP) *if it is conservative and both operations $\sqcap, \sqcup$ are commutative, i.e. $a \sqcap b = b \sqcap a$ and $a \sqcup b = b \sqcup a$ for all $a, b \in D$.*

- *An operation $\mathtt{Mj} : D^3 \to D$ is called a* majority operation *if for every tuple $(a, b, c) \in D^3$ with $|\{a, b, c\}| = 2$ operation $\mathtt{Mj}$ returns the unique majority element among $a, b, c$ (that occurs twice). An operation $\mathtt{Mn} : D^3 \to D$ is called a* minority operation *if for every tuple $(a, b, c) \in D^3$ with $|\{a, b, c\}| = 2$ operation $\mathtt{Mn}$ returns the unique minority element among $a, b, c$ (that occurs once).*

- *Triple $\langle \mathtt{Mj}_1, \mathtt{Mj}_2, \mathtt{Mn}_3 \rangle$ is called an* MJN *if it is conservative, $\mathtt{Mj}_1, \mathtt{Mj}_2$ are (possibly different) majority operations, and $\mathtt{Mn}_3$ is a minority operation.*

We say that $\langle \sqcap, \sqcup \rangle$ is a multimorphism of language $\Gamma$, or $\Gamma$ admits $\langle \sqcap, \sqcup \rangle$, if all cost functions $f \in \Gamma$ satisfy (1). Using a polynomial-time algorithm for minimising submodular functions, Cohen *et al.* have obtained the following result:

**Theorem 5** ([14]). *If a language $\Gamma$ admits an STP, then $\Gamma$ is tractable.*

The existence of an MJN multimorphism also leads to tractability. This was shown for a specific choice of an MJN by Cohen *et al.* [15].

Our tractability result, presented in the next section, will include both above-mentioned tractable classes as special cases.

**Expressibility** Finally, we define the important notion of expressibility, which captures the idea of introducing auxiliary variables in a VCSP instance and the possibility of minimising over these auxiliary variables. (For crisp languages, this is equivalent to *implementation* [19].)

**Definition 6.** *A cost function* $f : D^m \to \overline{\mathbb{Q}}_+$ *is* expressible *over a language* $\Gamma$ *if there exists an instance* $\mathcal{I} \in \mathsf{VCSP}(\Gamma)$ *with the set of nodes* $V = \{1, \ldots, m, m+1, \ldots, m+k\}$ *where* $k \geq 0$ *such that*

$$f(\boldsymbol{x}) = \min_{\boldsymbol{y} \in D^k} Cost_{\mathcal{I}}(\boldsymbol{x}, \boldsymbol{y}) \qquad \forall \boldsymbol{x} \in D^m$$

*We define* $\Gamma^*$ *to be the* expressive power *of* $\Gamma$*; that is, the set of all cost functions* $f$ *such that* $f$ *is expressible over* $\Gamma$.

The importance of expressibility is in the following result:

**Theorem 7** ([15])**.** *For any language* $\Gamma$, $\Gamma$ *is tractable iff* $\Gamma^*$ *is tractable.*

It is easy to observe and well known that any polymorphism (multimorphism) of $\Gamma$ is also a polymorphism (multimorphism) of $\Gamma^*$ [15].

## 3 Our results

In this section, we relate the complexity of a conservative language $\Gamma$ to properties of a certain graph $G_\Gamma$ associated with $\Gamma$.

Given a conservative language $\Gamma$, let $G_\Gamma = (P, E)$ be the graph with the set of nodes $P = \{(a, b) | a, b \in D, a \neq b\}$ and the set of edges $E$ defined as follows: there is an edge between $(a, b) \in P$ and $(a', b') \in P$ iff there exists binary cost function $f \in \Gamma^*$ such that

$$f(a, a') + f(b, b') > f(a, b') + f(b, a') , \quad (a, b'), (b, a') \in \mathrm{dom}\, f \qquad (3)$$

Note that $G_\Gamma$ may have self-loops. For node $p \in P$ we denote the self-loop by $\{p, p\}$. We say that edge $\{(a, b), (a', b')\} \in E$ is *soft* if there exists binary $f \in \Gamma^*$ satisfying (3) such that at least one of the assignments $(a, a'), (b, b')$ is in $\mathrm{dom}\, f$. Edges in $E$ that are not soft are called *hard*. For node $p = (a, b) \in P$ we denote $\bar{p} = (b, a) \in P$. Note, a somewhat similar graph (but not the same) was used by Takhanov [46] for languages $\Gamma$ containing crisp functions and finite unary cost functions.[1]

We denote $M \subseteq P$ to be the set of vertices $(a, b) \in P$ without self-loops, and $\overline{M} = P - M$ to be the complement of $M$. It follows from the definition that set $M$ is *symmetric*, i.e. $(a, b) \in M$ iff $(b, a) \in M$. We will write $\{a, b\} \in M$ to indicate that $(a, b) \in M$; this is consistent due to the symmetry of $M$. Similarly, we will write $\{a, b\} \in \overline{M}$ if $(a, b) \in \overline{M}$, and $\{a, b\} \in P$ if $(a, b) \in P$, i.e. $a, b \in D$ and $a \neq b$.

**Definition 8.** *Let* $\langle \sqcap, \sqcup \rangle$ *and* $\langle \mathtt{Mj}_1, \mathtt{Mj}_2, \mathtt{Mn}_3 \rangle$ *be binary and ternary operations respectively.*

- *Pair* $\langle \sqcap, \sqcup \rangle$ *is an* STP *on* $M$ *if* $\langle \sqcap, \sqcup \rangle$ *is conservative on* $P \cup \{\{a\} \mid a \in D\}$ *and commutative on* $M$.

---
[1]Roughly speaking, the graph structure in [46] was defined via a "min" polymorphism rather than a $\langle \min, \max \rangle$ multimorphism, so the property $\{p, q\} \in E \Rightarrow \{\bar{p}, \bar{q}\} \in E$ (that we prove for our graph in the next section) might not hold in Takhanov's case. Also, in [46] edges were not classified as being soft or hard.

- *Triple $\langle \mathtt{Mj}_1, \mathtt{Mj}_2, \mathtt{Mn}_3 \rangle$ is an MJN on $\overline{M}$ if it is conservative and for each triple $(a,b,c) \in D^3$ with $\{a,b,c\} = \{x,y\} \in \overline{M}$ operations $\mathtt{Mj}_1(a,b,c)$, $\mathtt{Mj}_2(a,b,c)$ return the unique majority element among $a,b,c$ (that occurs twice) and $\mathtt{Mn}_3(a,b,c)$ returns the remaining minority element.*

Our main results are given by the following three theorems.

**Theorem 9.** *Let $\Gamma$ be a conservative language.*

*(a) If $G_\Gamma$ has a soft self-loop then $\Gamma$ is NP-hard.*

*(b) If $G_\Gamma$ does not have soft self-loops then $\Gamma$ admits a pair $\langle \sqcup, \sqcap \rangle$ which is an STP on $M$ and satisfies additionally $a \sqcap b = a, a \sqcup b = b$ for $\{a,b\} \in \overline{M}$.*

**Theorem 10.** *Let $\Gamma$ be a conservative language. If $\Gamma$ does not admit an MJN on $\overline{M}$ then it is NP-hard.*

**Theorem 11.** *Suppose language $\Gamma$ admits an STP on $M$ and an MJN on $\overline{M}$, for some choice of symmetric $M \subseteq P$. Then $\Gamma$ is tractable.*

Theorems 9-11 give the dichotomy result for conservative languages:

**Corollary 12.** *If a conservative language $\Gamma$ admits an STP on $M$ and an MJN on $\overline{M}$ for some symmetric $M \subseteq P$ then $\Gamma$ is tractable. Otherwise $\Gamma$ is NP-hard.*

*Proof.* The first part follows from Theorem 11; let us show the second part. Suppose that the precondition of the corollary does not hold, then one of the following cases must be true (we assume below that $M$ is the set of nodes without self-loops in $G_\Gamma$):

- $G_\Gamma$ has a soft self-loop. Then $\Gamma$ is NP-hard by Theorem 9(a).

- $G_\Gamma$ does not have soft self-loops and $\Gamma$ does not admit an STP on $M$. This is a contradiction by Theorem 9(b).

- $G_\Gamma$ does not have soft self-loops and $\Gamma$ does not admit an MJN on $\overline{M}$. Then $\Gamma$ is NP-hard by Theorem 10.

□

In the finite-valued case, we get a simpler tractability criterion:

**Corollary 13.** *If a conservative finite-valued language $\Gamma$ admits an STP then $\Gamma$ is tractable. Otherwise $\Gamma$ is NP-hard.* [2]

*Proof.* Consider the graph $G_\Gamma$ associated with $\Gamma$. If $G_\Gamma$ contains a soft self-loop, then, by Theorem 9(a), $\Gamma$ is NP-hard. Suppose that $G_\Gamma$ does not contain soft self-loops. As $\Gamma$ is finite-valued, $G_\Gamma$ cannot have hard self-loops. Therefore, $\overline{M}$ is empty and $M = P$. By Theorem 9(b), $\Gamma$ admits an STP. The tractability then follows from Theorem 11. □

---

[2]It can be shown that if a finite-valued language admits an STP multimorphism, it also admits a submodularity multimorphism. This result is implicitly contained in [14]. Namely, after reducing the domains as in [14, Theorem 8.3], the STP might contain cycles. [14, Lemma 7.15] tells us that on cycles we have, in the finite-valued case, only unary cost functions. It follows that the cost functions admitting the STP must be submodular w.r.t. some total order [17].

This simplifies the tractability criterion in the finite-valued case (though we do not exploit this fact anywhere in the paper).

# 4 Proof preliminaries: strengthening the definition of conservativity

First, we show that we can strengthen the definition of conservative languages without loss of generality. More precisely, we prove in this section that it suffices to establish Theorems 9 and 10 under the following simplifying assumption:

**Assumption 1.** $\Gamma$ contains all general-valued unary cost functions $u : D \to \overline{\mathbb{Q}}_+$.

Let $\bar\Gamma$ be the language obtained from $\Gamma$ by adding all possible general-valued unary cost functions $u : D \to \overline{\mathbb{Q}}_+$. Note, $\bar\Gamma$ may be different from $\Gamma$ since $\Gamma$ is only guaranteed to have all possible $\{0,1\}$-valued unary cost functions.

**Proposition 14.** *(a) Graphs $G_\Gamma$ and $G_{\bar\Gamma}$ are the same: if $\{(a,b),(a',b')\}$ is a soft (hard) edge in $G_\Gamma$ then it is also a soft (hard) edge in $G_{\bar\Gamma}$, and vice versa. (b) If $\bar\Gamma$ is NP-hard then so is $\Gamma$.*

*Proof.* **Part (a)** One direction is trivial: if $\{(a,b),(a',b')\} \in G_\Gamma$ then $\{(a,b),(a',b')\} \in G_{\bar\Gamma}$, and if $\{(a,b),(a',b')\}$ is soft in $G_\Gamma$ then it is also soft in $G_{\bar\Gamma}$. For the other direction we need to show the following: (i) if $\{(a,b),(a',b')\}$ is an edge in $G_{\bar\Gamma}$ then it is also an edge in $G_\Gamma$, and (ii) if $\{(a,b),(a',b')\}$ is a soft edge in $G_{\bar\Gamma}$ then it is also soft in $G_\Gamma$.

Suppose that $\{(a,b),(a',b')\} \in G_{\bar\Gamma}$. Let $f \in (\bar\Gamma)^*$ be the corresponding binary function. If the edge $\{(a,b),(a',b')\}$ is soft in $G_{\bar\Gamma}$, then we choose $f$ according to the definition of the soft edge. We have

$$f(x,y) = \min_{\boldsymbol{z} \in D^{m-2}} g(x,y,\boldsymbol{z}) \qquad \forall x,y \in D$$

where $g : D^m \to \overline{\mathbb{Q}}_+$ is a sum of cost functions from $\bar\Gamma$. We can assume without loss of generality that all unary terms present in this sum are $\mathbb{Z} \cup \{\infty\}$-valued. Indeed, this can be ensured by multiplying $g$ by an appropriate integer $R$. (More precisely, unary terms $u : D \to \overline{\mathbb{Q}}_+$ in the sum are replaced with terms $R \cdot u \in \bar\Gamma$, and other terms $h$ in the sum are replaced by $R$ copies of $h$.)

Let $C$ be a sufficiently large finite integer constant (namely, $C > \max\{g(\boldsymbol{z}) \mid \boldsymbol{z} \in \mathrm{dom}\, g\}$), and let $g^C$ be the function obtained from $g$ as follows: we take every unary cost function $u : D \to \overline{\mathbb{Q}}_+$ present in $g$ and replace it with function $u^C(z) = \min\{u(z), C\}$. Clearly, $g^C \in \Gamma^*$. Define

$$f^C(x,y) = \min_{\boldsymbol{z} \in D^{m-2}} g^C(x,y,\boldsymbol{z}) \qquad \forall x,y \in D$$

then $f^C \in \Gamma^*$. It is easy to see that $f$ and $f^C$ have the following relationship: (i) if $f(x,y) < \infty$ then $f^C(x,y) = f(x,y) < C$; (ii) if $f(x,y) = \infty$ then $f^C(x,y) \geq C$. We have $f(a,a') + f(b,b') > f(a,b') + f(b,a')$ and $(a,b'),(b,a') \in \mathrm{dom}\, f$; this implies that $f^C(a,a') + f^C(b,b') > f^C(a,b') + f^C(b,a')$, and thus $\{p,q\} \in G_\Gamma$. If edge $\{p,q\}$ is soft in $G_{\bar\Gamma}$ then at least one of the assignments $(a,a'),(b,b')$ is in $\mathrm{dom}\, f$ (and thus in $\mathrm{dom}\, f^C$), and so $\{p,q\}$ is soft in $G_\Gamma$.

**Part (b)** Suppose that $\bar\Gamma$ is NP-hard, i.e. there exists a finite language $\bar\Gamma' \subseteq \bar\Gamma$ which is NP-hard. Let $\Gamma'$ be the language obtained from $\bar\Gamma'$ by first removing unary cost function $u : D \to \overline{\mathbb{Q}}_+$ present in $\bar\Gamma'$, and then adding all possible $\{0,1\}$-valued unary cost functions $u : D \to \{0,1\}$. Clearly, $\Gamma' \subseteq \Gamma$. We prove below that $\Gamma'$ is NP-hard using a reduction from $\bar\Gamma'$.

Let $R$ be a constant integer number such that multiplying unary cost functions from $\bar\Gamma'$ by $R$ gives $\mathbb{Z} \cup \{\infty\}$-valued functions. Also let $C_\circ$ be a sufficiently large finite integer constant, namely $C_\circ > \max\{R \cdot f(\boldsymbol{x}) \mid f \in \bar\Gamma', \boldsymbol{x} \in \mathrm{dom}\, f\}$. Now consider instance $\bar{\mathcal{I}}$ from $\bar\Gamma'$ with the cost function

$$f(\boldsymbol{x}) = \sum_{t \in T_1} u_t\left(x_{i(t,1)}\right) + \sum_{t \in T_*} f_t\left(x_{i(t,1)}, \ldots, x_{i(t,m_t)}\right)$$

where $T_1$ is the index set of unary cost functions and $T_*$ is index the set of cost functions of higher arities. Thus, $u_t \in \bar{\Gamma}'$ for $t \in T_1$ and $f_t \in \bar{\Gamma}'$ for $t \in T_*$. For each $t \in T_1$ we define unary cost function $u_t^C(z) = \min\{R \cdot u_t(z), C\}$ where $C = C_\circ \cdot (|T_1|+|T_*|)$. Note, we have $C > \max\{R \cdot f(\boldsymbol{x}) \,|\, \boldsymbol{x} \in \mathrm{dom}\, f\}$.

Let us define instance $\mathcal{I}$ with the cost function

$$f^C(\boldsymbol{x}) = \sum_{t \in T_1} u_t^C\left(x_{i(t,1)}\right) + \sum_{t \in T_*} R \cdot f_t\left(x_{i(t,1)}, \ldots, x_{i(t,m_t)}\right)$$

It can be viewed as an instance from $\Gamma'$. Indeed, $u_t^C$ can be represented as a sum of at most $C$ $\{0,1\}$-valued unary cost functions from $\Gamma'$, and the multiplication of $R$ and $f_t$ can be simulated by repeating the latter term $R$ times. Then $f_C$ contains at most $C|T_1| + R|T_*| = C_\circ(|T_1|+|T_*|)|T_1| + R|T_*|$ terms, so the size of instance $\mathcal{I}$ is bounded by a polynomial function of the size of $\bar{\mathcal{I}}$.

It is easy to see that $f$ and $f^C$ have the following relationship: (i) if $f(\boldsymbol{x}) < \infty$ then $f^C(\boldsymbol{x}) = R \cdot f(\boldsymbol{x}) < C$; (ii) if $f(\boldsymbol{x}) = \infty$ then $f^C(\boldsymbol{x}) \geq C$. Thus, solving $\mathcal{I}$ will also solve $\bar{\mathcal{I}}$. □

Proposition 14 shows that it suffices to prove Theorems 9 and 10 for language $\bar{\Gamma}$. Indeed, consider Theorem 9 for a conservative language $G$. If $G_\Gamma$ has a soft self-loop then by Proposition 14(a) so does $G_{\bar{\Gamma}}$. Theorem 9(a) for language $\bar{\Gamma}$ would imply that $\bar{\Gamma}$ is NP-hard, and therefore $\Gamma$ is also NP-hard by Proposition 14(b). If $G_\Gamma$ does not have soft self-loops then neither does $G_{\bar{\Gamma}}$. Theorem 9(b) for language $\bar{\Gamma}$ would imply that $\bar{\Gamma}$ admits the appropriate multimorphism $\langle \sqcup, \sqcap \rangle$ which is an STP on $M$. (Note, the definition of $M$ is the same for both $\Gamma$ and $\bar{\Gamma}$ by proposition 14(a).) Since $\Gamma \subseteq \bar{\Gamma}$, $\langle \sqcup, \sqcap \rangle$ is also a multimorphism of $\Gamma$.

A similar argumentation holds for Theorem 10. If $\bar{\Gamma}$ admits an MJN on $\overline{M}$ then so does $\Gamma$. If $\bar{\Gamma}$ does not admit an MJN on $\overline{M}$ then Theorem 10 for $\bar{\Gamma}$ and Proposition 14(b) would imply that $\Gamma$ is NP-hard.

In conclusion, from now on we will assume that language $\Gamma$ satisfies Assumption 1 when proving Theorems 9 and 10.

## 5 Proof of Theorem 9

In Section 5.1 we will first prove part (a). Then in Section 5.2 we will prove some properties of $G_\Gamma$ assuming that $G_\Gamma$ does not have self-loops. Using these properties, we will construct an STP on $M$ in Section 5.3.

### 5.1 NP-hard case

In this section we prove Theorem 9(a). From the assumption, there is a binary $f \in \Gamma^*$ such that $f(a,a) + f(b,b) > f(a,b) + f(b,a)$, and at least of the assignments $(a,a), (b,b)$ is in $\mathrm{dom}\, f$. First, let us assume that both $(a,a)$ and $(b,b)$ are in $\mathrm{dom}\, f$. Clearly, $g \in \Gamma^*$, where $g(x,y) = f(x,y) + f(y,x)$ has the following properties: $g(a,b) = g(b,a)$ and at least one of $\{g(a,a), g(b,b)\}$ is strictly bigger than $g(a,b)$. Let $\alpha = g(a,a)$ and $\beta = g(b,b)$. If $\alpha \neq \beta$, let $\alpha < \beta$ (the other case is analogous). Using unary cost functions with cost $(\beta - \alpha)/2$, we can construct $h \in \Gamma^*$ satisfying $h(a,a) = h(b,b) > h(a,b) = h(b,a)$. Now if $h(a,a) = h(b,b) = 1$ and $h(a,b) = h(b,a) = 0$, this would correspond to the Max-SAT problem with XOR clauses, which is NP-hard [41]. Since adding a constant to all cost functions and scaling all costs by a constant factor do not affect the difficulty of solving a VCSP instance, and $\Gamma$ is conservative, we can conclude that $\Gamma$ is intractable.

Without loss of generality, let us now assume that $(a,a) \in \mathrm{dom}\, f$ and $(b,b) \notin \mathrm{dom}\, f$. Using this function $f$ and unary cost functions, we can express function $g \in \Gamma^*$ with $g(a,a) = g(a,b) = g(b,a) = \alpha$ and $g(b,b) = \infty$, where $\alpha$ is a finite constant. Since adding a constant to $g$ does not affect the difficulty of solving a VCSP instance, we can assume without loss of generality that $\alpha = 0$. Using $g$ and unary cost functions, we can now encode the maximum independent set problem in graphs, a well-known NP-hard problem [25]: every vertex is represented by a variable with domain $\{a,b\}$ ($a$ represents not in the set, $b$ represents in the set); an edge between two vertices imposes a binary term between the corresponding two variables with cost function $g$. For every variable $x$, there is a unary term with cost function $h$ defined as $h(a) = 1$, $h(b) = 0$, and $h(c) = \infty$ for $D - \{a,b\}$. It is clear that minimising the number of variables assigned $a$ is the same as maximising the number of variables assigned $b$, thus finding a maximum independent set in the graph. □

## 5.2 Properties of graph $G_\Gamma$

From now on we assume that $E$ does not have soft self-loops. Our goal is to show that $\Gamma$ admits an STP on $M$.

In the lemma below, a *path* of length $k$ is a sequence of edges $\{p_0,p_1\}, \{p_1,p_2\}, \ldots, \{p_{k-1},p_k\}$, where $\{p_{i-1},p_i\} \in E$. Note that we allow edge repetitions. A path is *even* iff its length is even. A path is a *cycle* if $p_0 = p_k$. If $X \subseteq P$ then $(X, E[X])$ denotes the subgraph of $(P, E)$ induced by $X$.

**Lemma 15.** *Graph $G_\Gamma = (P, E)$ satisfies the following properties:*

(a) $\{p, q\} \in E$ *implies* $\{\bar{p}, \bar{q}\} \in E$ *and vice versa. The two edges are either both soft or both hard.*

(b) *Suppose that $\{p, q\} \in E$ and $\{q, r\} \in E$, then $\{p, \bar{r}\} \in E$. If at least one of the first two edges is soft then the third edge is also soft.*

(c) *For each $p \in P$, nodes $p$ and $\bar{p}$ are either both in $M$ or both in $\overline{M}$.*

(d) *There are no edges from $M$ to $\overline{M}$.*

(e) *Graph $(M, E[M])$ does not have odd cycles.*

(f) *If node $p$ is not isolated (i.e. it has at least one incident edge $\{p, q\} \in E$) then $\{p, \bar{p}\} \in E$.*

(g) *Nodes $p \in \overline{M}$ do not have incident soft edges.*

*Proof.* **(a)** Follows from the definition.
**(b)** Let $p = (a_1, b_1)$, $q = (a_2, b_2)$ and $r = (a_3, b_3)$. From the definition of the graph, let $f, g \in \Gamma^*$ be binary cost functions such that $(*)$ $f(a_1, a_2) + f(b_1, b_2) > f(a_1, b_2) + f(b_1, a_2)$ and $g(a_2, a_3) + g(b_2, b_3) > g(a_2, b_3) + g(b_2, a_3)$. Without loss of generality, we can assume that

$$\begin{aligned} f(a_1, a_2) &= \alpha, & f(a_1, b_2) &= f(b_1, a_2) = \gamma, & f(b_1, b_2) &= \alpha' \\ g(a_2, a_3) &= \beta, & g(a_2, b_3) &= f(b_2, a_3) = \gamma, & g(b_2, b_3) &= \beta' \end{aligned} \tag{4}$$

This can be achieved by replacing $f$ with $f'(x, y) = f(x, y) + f(y, x)$ and adding a constant, and similarly for $g$; condition $(*)$ and the complexity of $\Gamma$ are unaffected. From $(*)$ we get $\alpha + \alpha' > 2\gamma$; thus, by adding unary terms to $f$ we can ensure that $\alpha > \gamma$ and $\alpha' > \gamma$. Similarly, we can assume that $\beta > \gamma$ and $\beta' > \gamma$. (Note that $\gamma$ must be finite.)

Let $h(x, z) = \min_{z \in D}\{f(x, y) + u_{\{a_2, b_2\}}(y) + g(y, z)\}$, where $u_{\{a_2, b_2\}}(y) = 0$ if $y \in \{a_2, b_2\}$, and $u_{\{a_2, b_2\}}(y) = \infty$ otherwise. From the definition of $h$ and (4) we get $h(a_1, a_3) = h(b_1, b_3) = 2\gamma$ and $h(a_1, b_3) = \gamma + \min\{\alpha, \beta'\} > 2\gamma$, $h(b_1, a_3) = \gamma + \min\{\alpha', \beta\} > 2\gamma$. Therefore, $h(a_1, b_3) + h(b_1, a_3) > h(a_1, a_3) + h(b_1, b_3)$, and so $\{p, \bar{r}\} \in E$.

Now suppose that at least of one of the edges $\{p,q\}, \{q,r\}$ is soft, then we can assume that either $(\alpha,\alpha') \neq (\infty,\infty)$ or $(\beta,\beta') \neq (\infty,\infty)$. In each case either at least one of $h(a_1,b_3), h(b_1,a_3)$ is finite, and thus $\{p,\bar{r}\}$ is soft.

**(c)** Follows from (a).

**(d)** Suppose $\{p,q\} \in E$ and $q \in \overline{M}$. The latter fact implies $\{q,q\} \in E$, so by (b) we have $\{p,\bar{q}\} \in E$. From (a) we also get $\{q,\bar{p}\} \in E$. Applying (b) again gives $\{p,p\} \in E$. Thus $p \in \overline{M}$.

**(e)** We prove by induction on $k$ that $(M, E[M])$ does not have cycles of length $2k+1$. For $k = 0$ the claim is by assumption (nodes of $M$ do not have self-loops). Suppose it holds for $k \geq 0$, and suppose that $(M, E[M])$ has a cycle $\mathcal{P}, \{p,q\}, \{q,r\}, \{r,s\}$ of length $2k+3$ where $\mathcal{P}$ is path from $s \in M$ to $p \in M$ of length $2k$. Properties (b) and (a) give respectively $\{p,\bar{r}\} \in E$ and $\{\bar{r},\bar{s}\} \in E$. Applying (b) again gives $\{p,s'\} \in E$, therefore $(M, E[M])$ has a cycle $\mathcal{P}, \{p,s\}$ of length $2k+1$. This contradicts the induction hypothesis.

**(f)** Follows from (b).

**(g)** Suppose $p \in \overline{M}$ (implying $E$ has a hard self-loop $\{p,p\}$) and $\{p,q\}$ is a soft edge in $E$. Properties (b) and (a) give respectively $\{p,\bar{q}\} \in E$ and $\{\bar{q},\bar{p}\} \in E$, and furthermore both edges are soft. Applying (b) again gives that $\{p,p\} \in E$ and this edge is soft. This contradicts the assumption that $(P,E)$ does not have soft self-loops. $\square$

## 5.3 Constructing $\langle \sqcap, \sqcup \rangle$

In this section we complete the proof of Theorem 9 by constructing a pair of operations $\langle \sqcap, \sqcup \rangle$ for $\Gamma$ that behaves as an STP on $M$ and as a multi-projection (returning its two arguments in the same order) on $\overline{M}$.

**Lemma 16.** *There exists an assignment $\sigma : M \to \{-1,+1\}$ such that (i) $\sigma(p) = -\sigma(q)$ for all edges $\{p,q\} \in E$, and (ii) $\sigma(p) = -\sigma(\bar{p})$ for all $p \in M$.*

*Proof.* By Lemma 15(e) graph $(M, E[M])$ does not have odd cycles. Therefore, by Harary's Theorem, graph $(M, E[M])$ is bipartite and there exists an assignment $\sigma : M \to \{-1,+1\}$ that satisfies property (i). Let us modify this assignment as follows: for each isolated node $p \in M$ (i.e. node without incident edges) set $\sigma(p), \sigma(\bar{p})$ so that $\sigma(p) = -\sigma(\bar{p}) \in \{-1,+1\}$. (Note, if $p$ is isolated then by Lemma 15(a) so is $\bar{p}$). Clearly, property (i) still holds. Property (ii) holds for each node $p \in M$ as well: if $p$ is isolated then (ii) holds by construction, otherwise by Lemma 15(f) there exists edge $\{p,\bar{p}\} \in E$, and so (ii) follows from property (i). $\square$

Given assignment $\sigma$ constructed in Lemma 16, we now define operations $\sqcap, \sqcup : D^2 \to D$ as follows:

- $a \sqcap a = a \sqcup a = a$ for $a \in D$.

- If $(a,b) \in M$ then $a \sqcap b$ and $a \sqcup b$ are the unique elements of $D$ satisfying $\{a \sqcap b, a \sqcup b\} = \{a,b\}$ and $\sigma(a \sqcap b, a \sqcup b) = +1$.

- If $(a,b) \in \overline{M}$ then $a \sqcap b = a$ and $a \sqcup b = b$.

**Lemma 17.** *For any binary cost function $f \in \Gamma^*$ and any $\boldsymbol{x}, \boldsymbol{y} \in \mathrm{dom}\, f$ there holds*

$$f(\boldsymbol{x} \sqcap \boldsymbol{y}) + f(\boldsymbol{x} \sqcup \boldsymbol{y}) \leq f(\boldsymbol{x}) + f(\boldsymbol{y}) \tag{5}$$

*Proof.* Denote $(a, a') = \boldsymbol{x} \sqcap \boldsymbol{y}$ and $(b, b') = \boldsymbol{x} \sqcup \boldsymbol{y}$. We can assume without loss of generality that $\{\boldsymbol{x}, \boldsymbol{y}\} \ne \{(a, a'), (b, b')\}$, otherwise the claim is straightforward. It is easy to check that the assumption has two implications: (i) $a \ne b$ and $a' \ne b'$; (ii) $\{\boldsymbol{x}, \boldsymbol{y}\} = \{(a, b'), (b, a')\}$.

If $f(a, a') + f(b, b') = f(a, b') + f(b, a')$, then (5) holds trivially. If $f(a, a') + f(b, b') \ne f(a, b') + f(b, a')$, then $E$ contains at least one of the edges $\{(a, b), (a', b')\}$, $\{(a, b), (b', a')\}$. By Lemma 15(c) and Lemma 15(d), pairs $(a, b)$ and $(a', b')$ must either be both in $\overline{M}$ or both in $M$. In the former case (5) is a trivial equality from the definition of $\sqcap$ and $\sqcup$, so we assume the latter case.

The definition of $\sqcap, \sqcup$ and the fact that $(a, a') = \boldsymbol{x} \sqcap \boldsymbol{y}$ and $(b, b') = \boldsymbol{x} \sqcup \boldsymbol{y}$ imply that $\sigma(a, b) = \sigma(a', b') = +1$. Thus, set $E$ does not have edge $((a, b), (a', b'))$, and therefore

$$f(a, a') + f(b, b') \le f(a, b') + f(b, a')$$

which is equivalent to (5). □

In order to proceed, we introduce the following notation. Given a cost function $f$ of arity $m$, we denote by $V$ the set of variables corresponding to the arguments of $f$, with $|V| = m$. For two assignments $\boldsymbol{x}, \boldsymbol{y} \in D^m$ we denote $\Delta(\boldsymbol{x}, \boldsymbol{y}) = \{i \in V \mid x_i \ne y_i\}$ to be the set of variables on which $\boldsymbol{x}$ and $\boldsymbol{y}$ differ.

**Lemma 18.** *Condition* (5) *holds for any cost function* $f \in \Gamma^*$ *and assignments* $\boldsymbol{x}, \boldsymbol{y} \in \text{dom} f$ *with* $|\Delta(\boldsymbol{x}, \boldsymbol{y})| \le 2$.

*Proof.* If $|\Delta(\boldsymbol{x}, \boldsymbol{y})| \le 1$ then $\{\boldsymbol{x} \sqcap \boldsymbol{y}, \boldsymbol{x} \sqcup \boldsymbol{y}\} = \{\boldsymbol{x}, \boldsymbol{y}\}$, so the claim is trivial. We now prove it in the case $|\Delta(\boldsymbol{x}, \boldsymbol{y})| = 2$ using induction on $|V|$. The base case $|V| = 2$ follows from Lemma 17; suppose that $|V| \ge 3$. Choose $k \in V - \Delta(\boldsymbol{x}, \boldsymbol{y})$. For simplicity of notation, let us assume that $k$ corresponds to the first argument of $f$. Define cost function of $|V| - 1$ variables as

$$g(\boldsymbol{z}) = \min_{a \in D}\{u(a) + f(a, \boldsymbol{z})\} \qquad \forall \boldsymbol{z} \in D^{V - \{k\}} \tag{6}$$

where $u$ is the following unary cost function: $u(a) = 0$ if $a = x_k = y_k$, and $u(a) = \infty$ otherwise.

Let $\hat{\boldsymbol{x}}$ and $\hat{\boldsymbol{y}}$ be the restrictions of respectively $\boldsymbol{x}$ and $\boldsymbol{y}$ to $V - \{k\}$. Clearly, $g \in \Gamma^*$, $g(\hat{\boldsymbol{x}}) = f(\boldsymbol{x}) < \infty$ and $g(\hat{\boldsymbol{y}}) = f(\boldsymbol{y}) < \infty$. By the induction hypothesis

$$g(\hat{\boldsymbol{x}} \sqcap \hat{\boldsymbol{y}}) + g(\hat{\boldsymbol{x}} \sqcup \hat{\boldsymbol{y}}) \le g(\hat{\boldsymbol{x}}) + g(\hat{\boldsymbol{y}}) = f(\boldsymbol{x}) + f(\boldsymbol{y}) \tag{7}$$

This implies that $g(\hat{\boldsymbol{x}} \sqcap \hat{\boldsymbol{y}}) < \infty$, which is possible only if $g(\hat{\boldsymbol{x}} \sqcap \hat{\boldsymbol{y}}) = f(a, \hat{\boldsymbol{x}} \sqcap \hat{\boldsymbol{y}}) = f(\boldsymbol{x} \sqcap \boldsymbol{y})$ where $a = x_k = y_k$. Similarly, $g(\hat{\boldsymbol{x}} \sqcup \hat{\boldsymbol{y}}) = f(a, \hat{\boldsymbol{x}} \sqcup \hat{\boldsymbol{y}}) = f(\boldsymbol{x} \sqcup \boldsymbol{y})$. Thus, (7) is equivalent to (5). □

**Lemma 19.** *Condition* (5) *holds for any cost function* $f \in \Gamma^*$ *and any* $\boldsymbol{x}, \boldsymbol{y} \in \text{dom} f$.

*Proof.* We use induction on $|\Delta(\boldsymbol{x}, \boldsymbol{y})|$. The base case $|\Delta(\boldsymbol{x}, \boldsymbol{y})| \le 2$ follows from Lemma 18; suppose that $|\Delta(\boldsymbol{x}, \boldsymbol{y})| \ge 3$. Let us partition $\Delta(\boldsymbol{x}, \boldsymbol{y})$ into three sets $A, B, C$ as follows:

$$\begin{aligned}
A &= \{i \in \Delta(\boldsymbol{x}, \boldsymbol{y}) \mid (x_i, y_i) \in M, \ x_i = x_i \sqcap y_i, \ y_i = x_i \sqcup y_i\} \\
B &= \{i \in \Delta(\boldsymbol{x}, \boldsymbol{y}) \mid (x_i, y_i) \in M, \ x_i = x_i \sqcup y_i, \ y_i = x_i \sqcap y_i\} \\
C &= \{i \in \Delta(\boldsymbol{x}, \boldsymbol{y}) \mid (x_i, y_i) \in \overline{M}\}
\end{aligned}$$

Two cases are possible.

**Case 1** $|A \cup C| \ge 2$. Let us choose variable $k \in A \cup C$, and define assignments $\boldsymbol{x}', \boldsymbol{y}'$ as follows: $x'_i = y'_i = x_i = y_i$ if $x_i = y_i$, and for other variables

$$x'_i = \begin{cases} x_i & \text{if } i = k \\ y_i & \text{if } i \in (A \cup C) - \{k\} \\ x_i & \text{if } i \in B \end{cases} \qquad y'_i = \begin{cases} x_i & \text{if } i = k \\ y_i & \text{if } i \in (A \cup C) - \{k\} \\ y_i & \text{if } i \in B \end{cases}$$

It can be checked that

$$x \sqcap y' = x \sqcap y \qquad x \sqcup y' = x' \qquad x' \sqcap y = y' \qquad x' \sqcup y = x \sqcup y$$

Furthermore, $\Delta(x, y') = \Delta(x, y) - \{k\}$ and $\Delta(x', y) = \Delta(x, y) - ((A \cup C) - \{k\})$ so by the induction hypothesis

$$f(x \sqcap y) + f(x') \leq f(x) + f(y') \tag{8}$$

assuming that $y' \in \text{dom} f$, and

$$f(y') + f(x \sqcup y) \leq f(x') + f(y) \tag{9}$$

assuming that $x' \in \text{dom} f$. Two cases are possible:

- $y' \in \text{dom} f$. Inequality (8) implies that $x' \in \text{dom} f$. The claim then follows from summing (8) and (9).

- $y' \notin \text{dom} f$. Inequality (9) implies that $x' \notin \text{dom} f$. Assume for simplicity of notation that $k$ corresponds to the first argument of $f$. Define cost function of $|V| - 1$ variables

$$g(z) = \min_{a \in D}\{u(a) + f(a, z)\} \qquad \forall z \in D^{V - \{k\}}$$

where $u(a)$ is the following unary cost function: $u(x_k) = 0$, $u(y_k) = C$ and $u(a) = \infty$ for $a \in D - \{x_k, y_k\}$. Here $C$ is a sufficiently large finite constant, namely $C > f(x) + f(y)$.

Let $\hat{x}, \hat{y}, \hat{x}', \hat{y}'$ be restrictions of respectively $x, y, x', y'$ to $V - \{k\}$. Clearly, $g \in \Gamma^*$ and

$$\begin{aligned} g(\hat{y}) = g(\hat{y}') &= u(y_k) + f(y_k, \hat{y}) = f(y) + C \qquad \text{(since } (x_k, \hat{y}) = y' \notin \text{dom} f) \\ g(\hat{x}) &= f(x_k, \hat{x}) = f(x) \end{aligned}$$

By the induction hypothesis

$$g(\hat{x} \sqcap \hat{y}) + g(\hat{x} \sqcup \hat{y}) \leq g(\hat{x}) + g(\hat{y}) = f(x) + f(y) + C \tag{10}$$

We have $g(\hat{x} \sqcup \hat{y}) < \infty$, so we must have either $g(\hat{x} \sqcup \hat{y}) = f(x_k, \hat{x} \sqcup \hat{y})$ or $g(\hat{x} \sqcup \hat{y}) = f(y_k, \hat{x} \sqcup \hat{y}) + C = f(x \sqcup y) + C$. The former case is impossible since $(x_k, \hat{x} \sqcup \hat{y}) = x' \notin \text{dom} f$, so $g(\hat{x} \sqcup \hat{y}) = f(x \sqcup y) + C$. Combining it with (10) gives

$$g(\hat{x} \sqcap \hat{y}) + f(x \sqcup y) \leq f(x) + f(y) \tag{11}$$

This implies that $g(\hat{x} \sqcap \hat{y}) < C$, so we must have $g(\hat{x} \sqcap \hat{y}) = f(x_k, \hat{x} \sqcap \hat{y}) = f(x \sqcap y)$. Thus, (11) is equivalent to (5).

**Case 2** $|B| \geq 2$. Let us choose variable $k \in B$, and define assignments $x', y'$ as follows: $x'_i = y'_i = x_i = y_i$ if $x_i = y_i$, and for other variables

$$x'_i = \begin{cases} y_i & \text{if } i = k \\ x_i & \text{if } i \in A \cup C \\ x_i & \text{if } i \in B - \{k\} \end{cases} \qquad y'_i = \begin{cases} y_i & \text{if } i = k \\ y_i & \text{if } i \in A \cup C \\ x_i & \text{if } i \in B - \{k\} \end{cases}$$

It can be checked that

$$x' \sqcap y = x \sqcap y \qquad x' \sqcup y = y' \qquad x \sqcap y' = x' \qquad x \sqcup y' = x \sqcup y$$

Furthermore, $\Delta(\boldsymbol{x}', \boldsymbol{y}) = \Delta(\boldsymbol{x}, \boldsymbol{y}) - \{k\}$ and $\Delta(\boldsymbol{x}, \boldsymbol{y}') = \Delta(\boldsymbol{x}, \boldsymbol{y}) - (B - \{k\})$ so by the induction hypothesis

$$f(\boldsymbol{x} \sqcap \boldsymbol{y}) + f(\boldsymbol{y}') \leq f(\boldsymbol{x}') + f(\boldsymbol{y}) \qquad (12)$$

assuming that $\boldsymbol{x}' \in \mathrm{dom}\, f$, and

$$f(\boldsymbol{x}') + f(\boldsymbol{x} \sqcup \boldsymbol{y}) \leq f(\boldsymbol{x}) + f(\boldsymbol{y}') \qquad (13)$$

assuming that $\boldsymbol{y}' \in \mathrm{dom}\, f$. Two cases are possible:

- $\boldsymbol{x}' \in \mathrm{dom}\, f$. Inequality (12) implies that $\boldsymbol{y}' \in \mathrm{dom}\, f$. The claim then follows from summing (12) and (13).

- $\boldsymbol{x}' \notin \mathrm{dom}\, f$. Inequality (13) implies that $\boldsymbol{y}' \notin \mathrm{dom}\, f$. Assume for simplicity of notation that $k$ corresponds to the first argument of $f$. Define function of $|V| - 1$ variables

$$g(\boldsymbol{z}) = \min_{a \in D}\{u(a) + f(a, \boldsymbol{z})\} \qquad \forall \boldsymbol{z} \in D^{V-\{k\}}$$

where $u(a)$ is the following unary term: $u(y_k) = 0$, $u(x_k) = C$ and $u(a) = \infty$ for $a \in D - \{x_k, y_k\}$. Here $C$ is a sufficiently large finite constant, namely $C > f(\boldsymbol{x}) + f(\boldsymbol{y})$.

Let $\hat{\boldsymbol{x}}, \hat{\boldsymbol{y}}, \hat{\boldsymbol{x}}', \hat{\boldsymbol{y}}'$ be restrictions of respectively $\boldsymbol{x}, \boldsymbol{y}, \boldsymbol{x}', \boldsymbol{y}'$ to $V - \{k\}$. Clearly, $g \in \Gamma^*$ and

$$\begin{aligned} g(\hat{\boldsymbol{x}}) = g(\hat{\boldsymbol{x}}') &= u(x_k) + f(x_k, \hat{\boldsymbol{x}}) = f(\boldsymbol{x}) + C \qquad \text{(since } (y_k, \hat{\boldsymbol{x}}) = \boldsymbol{x}' \notin \mathrm{dom}\, f) \\ g(\hat{\boldsymbol{y}}) &= f(y_k, \hat{\boldsymbol{y}}) = f(\boldsymbol{y}) \end{aligned}$$

By the induction hypothesis

$$g(\hat{\boldsymbol{x}} \sqcap \hat{\boldsymbol{y}}) + g(\hat{\boldsymbol{x}} \sqcup \hat{\boldsymbol{y}}) \leq g(\hat{\boldsymbol{x}}) + g(\hat{\boldsymbol{y}}) = f(\boldsymbol{x}) + f(\boldsymbol{y}) + C \qquad (14)$$

We have $g(\hat{\boldsymbol{x}} \sqcup \hat{\boldsymbol{y}}) < \infty$, so we must have either $g(\hat{\boldsymbol{x}} \sqcup \hat{\boldsymbol{y}}) = f(y_k, \hat{\boldsymbol{x}} \sqcup \hat{\boldsymbol{y}})$ or $g(\hat{\boldsymbol{x}} \sqcup \hat{\boldsymbol{y}}) = f(x_k, \hat{\boldsymbol{x}} \sqcup \hat{\boldsymbol{y}}) + C = f(\boldsymbol{x} \sqcup \boldsymbol{y}) + C$. The former case is impossible since $(y_k, \hat{\boldsymbol{x}} \sqcup \hat{\boldsymbol{y}}) = \boldsymbol{y}' \notin \mathrm{dom}\, f$, so $g(\hat{\boldsymbol{x}} \sqcup \hat{\boldsymbol{y}}) = f(\boldsymbol{x} \sqcup \boldsymbol{y}) + C$. Combining it with (14) gives

$$g(\hat{\boldsymbol{x}} \sqcap \hat{\boldsymbol{y}}) + f(\boldsymbol{x} \sqcup \boldsymbol{y}) \leq f(\boldsymbol{x}) + f(\boldsymbol{y}) \qquad (15)$$

This implies that $g(\hat{\boldsymbol{x}} \sqcap \hat{\boldsymbol{y}}) < C$, so we must have $g(\hat{\boldsymbol{x}} \sqcap \hat{\boldsymbol{y}}) = f(y_k, \hat{\boldsymbol{x}} \sqcap \hat{\boldsymbol{y}}) = f(\boldsymbol{x} \sqcap \boldsymbol{y})$. Thus, (15) is equivalent to (5).

□

## 6 Proof of Theorem 10

For a language $\Gamma$ let $Feas(\Gamma)$ be the language obtained from $\Gamma$ by converting all finite values of $f$ to 0 for all $f \in \Gamma$, and let $MinHom(\Gamma)$ be the language obtained from $Feas(\Gamma)$ by adding all possible integer-valued unary cost functions $u : D \to \mathbb{Z}_+$. Note, $MinHom(\Gamma)$ corresponds to the *minimum-cost homomorphism* problem introduced in [28] and recently studied in [46]. We will need the following fact which is a simple corollary of results of Takhanov [46].

**Theorem 20.** *(a) If $MinHom(\Gamma)$ does not admit a majority polymorphism then $MinHom(\Gamma)$ is NP-hard. (b) If $MinHom(\Gamma)$ is NP-hard then so is $\Gamma$.*

*Proof.*

**Part (a)** Takhanov has studied crisp languages including additionally all integer-valued unary cost functions [46]. For such a language $\Gamma$, he considers the functional clone of all polymorphisms of $\Gamma$, denoted by $F$, and a certain graph denoted by $T_F$. Takhanov's Theorem 3.3, Theorem 3.4, and Theorem 5.3 give the following:

- If $F$ does not satisfy the necessary local conditions or $T_F$ is not bipartite then $F$ is NP-hard.

- If $F$ satisfies the necessary local conditions and $T_F$ is bipartite then $F$ contains a majority operation.

This implies part (a).

**Part (b)** Let $MinHom(\Gamma)' \subseteq MinHom(\Gamma)$ be a finite language with costs in $\mathbb{Z}_+ \cup \{\infty\}$ which is NP-hard. Denote $MinHom(\Gamma)'_1$ and $MinHom(\Gamma)'_*$ to be the subsets of $MinHom(\Gamma)'$ of arity $m = 1$ and $m \geq 2$ respectively. The definition of $MinHom(\Gamma)$ implies that for every $f \in MinHom(\Gamma)'_*$ there exists function $f^\circ \in \Gamma$ such that $f(\boldsymbol{x}) = 0$ if $f^\circ(\boldsymbol{x}) < \infty$, and $f(\boldsymbol{x}) = \infty$ if $f^\circ(\boldsymbol{x}) = \infty$. Denote $C = \max\{f^\circ(\boldsymbol{x}) \mid f \in MinHom(\Gamma)'_*, \boldsymbol{x} \in \operatorname{dom} f^\circ\} + 1$. Construct language $\Gamma'$ as follows:

$$\Gamma' = \{u^C \mid u \in MinHom(\Gamma)'_1\} \ \cup \ \{f^\circ \mid f \in MinHom(\Gamma)'_*\}$$

where function $u^C$ is defined by $u^C(z) = C \cdot u(z)$. Clearly, $\Gamma' \subseteq \Gamma$. We prove below that $\Gamma'$ is NP-hard using a reduction from $MinHom(\Gamma)'$.

Let $\hat{\mathcal{I}}$ be an instance from $MinHom(\Gamma)'$ with the cost function

$$f(\boldsymbol{x}) = \sum_{t \in T_1} u_t\left(x_{i(t,1)}\right) + \sum_{t \in T_*} f_t\left(x_{i(t,1)}, \ldots, x_{i(t,m_t)}\right)$$

where $T_1$ is the index set of unary cost functions and $T_*$ is the index set of cost functions of higher arities. Note, $u_t \in MinHom(\Gamma)'_1$ for $t \in T_1$ and $f_t \in MinHom(\Gamma)'_*$ for $t \in T_*$. Now define instance $\mathcal{I}$ with the cost function

$$f^C(\boldsymbol{x}) = \sum_{t \in T_1} N \cdot u_t^C\left(x_{i(t,1)}\right) + \sum_{t \in T_*} f_t^\circ\left(x_{i(t,1)}, \ldots, x_{i(t,m_t)}\right)$$

where $N = |T_*|$. It can be viewed as an instance from $\Gamma'$, if we simulate multiplication of $N$ and $u_t^C$ by repeating the latter term $N$ times; the size of the expression grows only polynomially. For any $\boldsymbol{x} \in \operatorname{dom} f$ we have

$$f^C(\boldsymbol{x}) \geq \sum_{t \in T_1} N \cdot u_t^C\left(x_{i(t,1)}\right) = NC \cdot f(\boldsymbol{x})$$

$$f^C(\boldsymbol{x}) < \sum_{t \in T_1} N \cdot u_t^C\left(x_{i(t,1)}\right) + \sum_{t \in T_*} C = NC \cdot (f(\boldsymbol{x}) + 1)$$

Furthermore, $f(\boldsymbol{x}) = \infty$ iff $f^C(\boldsymbol{x}) = \infty$. Function $f$ have values in $\mathbb{Z}_+ \cup \{\infty\}$, therefore solving $\mathcal{I}$ will also solve $\hat{\mathcal{I}}$. □

Suppose that $\Gamma$ does not admit a majority polymorphism. Clearly, this implies that $MinHom(\Gamma)$ also does not admit a majority polymorphism. By Theorem 20, $\Gamma$ is NP-hard, and so Theorem 10 holds in this case. Hence without loss of generality we can assume:

**Assumption 2.** $\Gamma$ admits a majority polymorphism.

By Theorem 9(a), if $G_\Gamma$ has a soft self-loop then $\Gamma$ is NP-hard. Hence without loss of generality we can assume:

**Assumption 3.** $G_\Gamma$ does not have soft self-loops.

To prove Theorem 10, we need to show the existence of an MJN multimorphism on $\overline{M}$ under assumptions 1-3. We denote by $\langle \sqcap, \sqcup \rangle$ an STP multimorphism on $M$ with the properties given in Theorem 9(b).

## 6.1 Constructing $\langle \mathtt{Mj}_1, \mathtt{Mj}_2, \mathtt{Mn}_3 \rangle$

Let us introduce function $\mu$ which maps every set $\{a,b,c\} \subseteq D$ with $|\{a,b,c\}| = 3$ to a subset of $\{a,b,c\}$. This subset is defined as follows: $c \in \mu(\{a,b,c\})$ iff there exists binary function $f \in \Gamma^*$ and a pair $(a', b') \in \overline{M}$ such that
$$\mathtt{dom}\, f = \{(a, a'), (b, a'), (c, b')\}$$

**Lemma 21.** *Set $\mu(\{a,b,c\})$ contains at most one label. Furthermore, if $\mu(\{a,b,c\}) = \{c\}$ then $(a,c) \in \overline{M}$ and $(b,c) \in \overline{M}$.*

*Proof.* Suppose that $a, c \in \mu(\{a,b,c\})$ where $a \neq c$, then there exist binary functions $f, g \in \Gamma^*$ and pairs $(a', b'), (a'', b'') \in \overline{M}$ such that
$$\mathtt{dom}\, f = \{(a', a), (b', b), (b', c)\} \qquad \mathtt{dom}\, g = \{(a, a''), (b, a''), (c, b'')\}$$

Consider function
$$h(x', x'') = \min_{x \in D}\{f(x', x) + g(x, x'')\} \tag{16}$$

Clearly, $\mathtt{dom}\, h = \{(a', a''), (b', a''), (b', b'')\}$, so $(a', b') \in \overline{M}$ has an incident soft edge in $G_\Gamma$ - a contradiction.

This second claim of the lemma follows from Lemma 15(d). □

For convenience, we define $\mu(\{a,b,c\}) = \varnothing$ if $|\{a,b,c\}| \leq 2$. We are now ready to construct operation $\mathtt{MJN} = \langle \mathtt{Mj}_1, \mathtt{Mj}_2, \mathtt{Mn}_3 \rangle$. Given a tuple $(a,b,c) \in D^3$, we define

$$\mathtt{MJN}(a,b,c) = \begin{cases} (x, x, y) & \text{if } \{\!\{a,b,c\}\!\} = \{\!\{x,x,y\}\!\}, \{x,y\} \in \overline{M} & (17a) \\ (b \sqcap c, b \sqcup c, a) & \text{if } \mu(\{a,b,c\}) = \{a\} & (17b) \\ (a \sqcap c, a \sqcup c, b) & \text{if } \mu(\{a,b,c\}) = \{b\} & (17c) \\ (a \sqcap b, a \sqcup b, c) & \text{in any other case} & (17d) \end{cases}$$

where $\{\!\{\ldots\}\!\}$ denotes a *multiset*, i.e. elements' multiplicities are taken into account.

**Theorem 22.** *If $f \in \Gamma^*$ and $\boldsymbol{x}, \boldsymbol{y}, \boldsymbol{z} \in \mathtt{dom}\, f$ then*
$$f(\mathtt{Mj}_1(\boldsymbol{x}, \boldsymbol{y}, \boldsymbol{z})) + f(\mathtt{Mj}_2(\boldsymbol{x}, \boldsymbol{y}, \boldsymbol{z})) + f(\mathtt{Mn}_3(\boldsymbol{x}, \boldsymbol{y}, \boldsymbol{z})) \leq f(\boldsymbol{x}) + f(\boldsymbol{y}) + f(\boldsymbol{z}) \tag{18}$$

The remainder of Section 6 is devoted to the proof of this statement.

## 6.2 Proof of Theorem 22: preliminaries

We say that an instance $(f, \boldsymbol{x}, \boldsymbol{y}, \boldsymbol{z})$ is *valid* if $f \in \Gamma^*$ and $\boldsymbol{x}, \boldsymbol{y}, \boldsymbol{z} \in \text{dom} f$. It is *satisfiable* if (18) holds, and *unsatisfiable* otherwise. For a triple $\boldsymbol{x}, \boldsymbol{y}, \boldsymbol{z} \in D^V$ denote $\delta(\boldsymbol{x}, \boldsymbol{y}, \boldsymbol{z}) = \sum_{i \in V} |\{x_i, y_i, z_i\}|$, $\Delta(\boldsymbol{x}, \boldsymbol{y}, \boldsymbol{z}) = \{i \in V \mid x_i \neq y_i\}$ and $\Delta^M(\boldsymbol{x}, \boldsymbol{y}, \boldsymbol{z}) = \{i \in \Delta(\boldsymbol{x}, \boldsymbol{y}, \boldsymbol{z}) \mid \{x_i, y_i, z_i\} = \{a, b\} \in M\}$.

Suppose that an unsatisfiable instance exists. From now on we assume that $(f, \boldsymbol{x}, \boldsymbol{y}, \boldsymbol{z})$ is a lowest unsatisfiable instance with respect to the partial order $\preceq$ defined as the lexicographical order with components

$$( \ \delta(\boldsymbol{x}, \boldsymbol{y}, \boldsymbol{z}), \quad |\Delta(\boldsymbol{x}, \boldsymbol{y}, \boldsymbol{z})|, \quad |\Delta^M(\boldsymbol{x}, \boldsymbol{y}, \boldsymbol{z})|, \quad |\{i \in V \mid \mu(\{x_i, y_i, z_i\}) = \{x_i\}\}| \ ) \tag{19}$$

(the first component is more significant). We denote $\delta_{\min} = \delta(\boldsymbol{x}, \boldsymbol{y}, \boldsymbol{z})$. Thus, we have

**Assumption 4.** *All valid instances $(f, \boldsymbol{x}', \boldsymbol{y}', \boldsymbol{z}')$ with $(\boldsymbol{x}', \boldsymbol{y}', \boldsymbol{z}') \prec (\boldsymbol{x}, \boldsymbol{y}, \boldsymbol{z})$ (and in particular with $\delta(\boldsymbol{x}', \boldsymbol{y}', \boldsymbol{z}') < \delta_{\min}$) are satisfiable, while the instance $(f, \boldsymbol{x}, \boldsymbol{y}, \boldsymbol{z})$ is unsatisfiable.*

We will assume without loss of generality that for any $\boldsymbol{u} \in \text{dom} f$ there holds $u_i \in \{x_i, y_i, z_i\}$ for all $i \in V$. Indeed, this can be achieved by adding unary cost functions $g_i(u_i)$ to $f$ with $\text{dom} g_i = \{x_i, y_i, z_i\}$; this does not affect the satisfiability of $(f, \boldsymbol{x}, \boldsymbol{y}, \boldsymbol{z})$.

The following cases can be easily eliminated:

**Proposition 23.** *The following cases are impossible: (a) $|V| = 1$; (b) $|\{x_i, y_i, z_i\}| = 1$ for some $i \in V$.*

*Proof.* If $|V| = 1$ then (18) is a trivial equality contradicting to the choice of $(f, \boldsymbol{x}, \boldsymbol{y}, \boldsymbol{z})$. Suppose that $x_i = y_i = z_i = a$, $i \in V$. Consider function

$$g(\boldsymbol{u}) = \min_{d \in D} f(d, \boldsymbol{u}) \qquad \forall \boldsymbol{u} \in D^{\hat{V}}$$

where $\hat{V} = V - \{i\}$ and we assumed for simplicity of notation that $i$ corresponds to the first argument of $f$. For an assignment $\boldsymbol{w} \in V$ we denote $\hat{\boldsymbol{w}}$ to be the restriction of $\boldsymbol{w}$ to $\hat{V}$. Clearly, $g \in \Gamma^*$, $g(\hat{\boldsymbol{x}}) = f(\boldsymbol{x})$, $g(\hat{\boldsymbol{y}}) = f(\boldsymbol{y})$, $g(\hat{\boldsymbol{y}}) = f(\boldsymbol{y})$ and $(\hat{\boldsymbol{x}}, \hat{\boldsymbol{y}}, \hat{\boldsymbol{z}}) \prec (\boldsymbol{x}, \boldsymbol{y}, \boldsymbol{z})$, so Assumption 4 gives

$$g(\text{Mj}_1(\hat{\boldsymbol{x}}, \hat{\boldsymbol{y}}, \hat{\boldsymbol{z}})) + g(\text{Mj}_2(\hat{\boldsymbol{x}}, \hat{\boldsymbol{y}}, \hat{\boldsymbol{z}})) + g(\text{Mn}_3(\hat{\boldsymbol{x}}, \hat{\boldsymbol{y}}, \hat{\boldsymbol{z}})) \leq g(\hat{\boldsymbol{x}}) + g(\hat{\boldsymbol{y}}) + g(\hat{\boldsymbol{z}}) = f(\boldsymbol{x}) + f(\boldsymbol{y}) + f(\boldsymbol{z})$$

This implies that $\text{Mj}_1(\hat{\boldsymbol{x}}, \hat{\boldsymbol{y}}, \hat{\boldsymbol{z}}) \in \text{dom} g$ and thus $g(\text{Mj}_1(\hat{\boldsymbol{x}}, \hat{\boldsymbol{y}}, \hat{\boldsymbol{z}})) = f(a, \text{Mj}_1(\hat{\boldsymbol{x}}, \hat{\boldsymbol{y}}, \hat{\boldsymbol{z}})) = f(\text{Mj}_1(\boldsymbol{x}, \boldsymbol{y}, \boldsymbol{z}))$. Similarly, $g(\text{Mj}_2(\hat{\boldsymbol{x}}, \hat{\boldsymbol{y}}, \hat{\boldsymbol{z}})) = f(\text{Mj}_2(\boldsymbol{x}, \boldsymbol{y}, \boldsymbol{z}))$ and $g(\text{Mn}_3(\hat{\boldsymbol{x}}, \hat{\boldsymbol{y}}, \hat{\boldsymbol{z}})) = f(\text{Mn}_3(\boldsymbol{x}, \boldsymbol{y}, \boldsymbol{z}))$, so the inequality above is equivalent to (18). □

It is also easy to show the following fact.

**Proposition 24.** *There exists node $i \in V$ for which operation $\text{MJN}(x_i, y_i, z_i)$ is defined by equation (17a), (17b) or (17c), i.e. either $\{x_i, y_i, z_i\} = \{a, b\} \in \overline{M}$, $\mu(\{x_i, y_i, z_i\}) = \{x_i\}$, or $\mu(\{x_i, y_i, z_i\}) = \{y_i\}$.*

*Proof.* If such a node does not exist then $\text{MJN}(x_i, y_i, z_i)$ is defined by equation (17d) for all nodes $i \in V$, i.e. $\text{MJN}(\boldsymbol{x}, \boldsymbol{y}, \boldsymbol{z}) = (\boldsymbol{x} \sqcap \boldsymbol{y}, \boldsymbol{x} \sqcup \boldsymbol{y}, \boldsymbol{z})$. The fact that $\langle \sqcap, \sqcup \rangle$ is a multimorphism of $f$ then implies inequality (18), contradicting to the choice of $(f, \boldsymbol{x}, \boldsymbol{y}, \boldsymbol{z})$. □

In the next section we show that case (17a) is impossible, while the remaining two cases (17b), (17c) are analysed in section 6.4.

The following equalities are easy to verify; they will be useful for verifying various identities:

$$\alpha \sqcap (\alpha \sqcup \beta) = \alpha \sqcap (\beta \sqcup \alpha) = (\alpha \sqcap \beta) \sqcup \alpha = (\beta \sqcap \alpha) \sqcup \alpha = \alpha \qquad \forall \alpha, \beta \in D \tag{20a}$$

$$\text{MJN}(\alpha, \alpha, \beta) = (\alpha, \alpha, \beta) \qquad \forall \alpha, \beta \in D \tag{20b}$$

$$\{\{\text{Mj}_1(\alpha, \beta, \gamma), \text{Mj}_2(\alpha, \beta, \gamma), \text{Mn}_3(\alpha, \beta, \gamma)\}\} = \{\{\alpha, \beta, \gamma\}\} \qquad \forall \alpha, \beta, \gamma \in D \tag{20c}$$

## 6.3 Eliminating case (17a)

We will need the following result.

**Lemma 25.** *Suppose that $i \in V$ is a node with $\{\{x_i, y_i, z_i\}\} = \{\{a, b, b\}\}$ where $\{a, b\} \in \overline{M}$. Let $u \in \{x, y, z\}$ be the labelling with $u_i = a$, and let $u'$ be the labelling obtained from $u$ by setting $u'_i = b$. Then $u' \in \text{dom} f$.*

*Proof.* Assume that $u = x$ (the cases $u = y$ and $y = z$ will be entirely analogous). Accordingly, we denote $x' = u'$. By Assumption 2, $f$ admits a majority polymorphism. This implies [1] that $f$ is *decomposable into unary and binary relations*, i.e. there holds

$$u \in \text{dom} f \quad \Leftrightarrow \quad [u_i \in \text{dom} \rho_i \; \forall i \in V \text{ and } (u_i, u_j) \in \text{dom} \rho_{ij} \; \forall i, j \in V, i \neq j]$$

where unary functions $\rho_i \in \Gamma^*$ for $i \in V$ and binary functions $\rho_{ij} \in \Gamma^*$ for distinct $i, j \in V$ are defined as

$$\rho_i(a_i) = \min\{f(u) \mid u_i = a_i\} \qquad \forall a_i \in D$$
$$\rho_{ij}(a_i, a_j) = \min\{f(u) \mid (u_i, u_j) = (a_i, a_j)\} \qquad \forall (a_i, a_j) \in D^2$$

Suppose that $x' \notin \text{dom} f$, then there exists node $j \in V - \{i\}$ such that $(x'_i, x'_j) = (b, x_j) \notin \text{dom} \rho_{ij}$. We must have $(a, x_j), (b, y_j), (b, z_j) \in \text{dom} \rho_{ij}$ since $x, y, z \in \text{dom} f$. This implies, in particular, that $y_j \neq x_j$ and $z_j \neq x_j$. Furthermore, $(a, y_i), (a, z_i) \notin \text{dom} \rho_{ij}$, otherwise pair $(a, b) \in \overline{M}$ would have an incident soft edge in $G_\Gamma$. Two cases are possible:

- $y_j = z_j$. The edge $\{(a, b), (y_j, x_j)\}$ belongs to $G_\Gamma$, therefore $(x_j, y_j) \in \overline{M}$.
- $y_j \neq z_j$. We have $\text{dom} \rho_{ij} = \{(a, x_j), (b, y_j), (b, z_j)\}$, therefore $\mu(\{x_j, y_j, z_j\}) = \{x_j\}$.

In each case $\text{Mj}_1(x_j, y_j, z_j) \neq x_j$, $\text{Mj}_2(x_j, y_j, z_j) \neq x_j$ and $\text{Mn}_3(x_j, y_j, z_j) = x_j$. Now let us "minimise out" variable $x_i$, i.e. define function

$$g(u) = \min_{d \in D} f(d, u) \qquad \forall u \in D^{\hat{V}} \tag{21}$$

where $\hat{V} = V - \{i\}$ and we assumed that $i$ corresponds to the first argument of $f$. For an assignment $u \in V$ we denote $\hat{u}$ to be the restriction of $u$ to $\hat{V}$. Due to the presence of relation $\rho_{ij}$ we have

$$g(\hat{x}) = f(x) \qquad g(\text{Mj}_1(\hat{x}, \hat{y}, \hat{z})) = f(\text{Mj}_1(x, y, z))$$
$$g(\hat{y}) = f(y) \qquad g(\text{Mj}_2(\hat{x}, \hat{y}, \hat{z})) = f(\text{Mj}_2(x, y, z))$$
$$g(\hat{z}) = f(z) \qquad g(\text{Mn}_3(\hat{x}, \hat{y}, \hat{z})) = f(\text{Mn}_3(x, y, z))$$

Since $\delta(\hat{x}, \hat{y}, \hat{z}) < \delta(x, y, z)$, Assumption 4 gives

$$g(\text{Mj}_1(\hat{x}, \hat{y}, \hat{z})) + g(\text{Mj}_2(\hat{x}, \hat{y}, \hat{z})) + g(\text{Mn}_3(\hat{x}, \hat{y}, \hat{z})) \leq f(x) + f(y) + f(z)$$

which is equivalent to (18). □

Let us denote

$$V^M = \{i \in V \mid \{x_i, y_i, z_i\} = \{a, b\} \in M\}$$
$$V^{\overline{M}} = \{i \in V \mid \{x_i, y_i, z_i\} = \{a, b\} \in \overline{M}\}$$
$$V_1^{\overline{M}} = \{i \in V^{\overline{M}} \mid (x_i, y_i, z_i) = (a, b, b)\} \subseteq \Delta(x, y, z)$$
$$V_2^{\overline{M}} = \{i \in V^{\overline{M}} \mid (x_i, y_i, z_i) = (b, a, b)\} \subseteq \Delta(x, y, z)$$
$$V_3^{\overline{M}} = \{i \in V^{\overline{M}} \mid (x_i, y_i, z_i) = (b, b, a)\}$$

We need to show that $V^{\overline{M}}$ is empty.

**Proposition 26.** *Suppose that $i \in V^{\overline{M}}$.*

(a) *If $(x_i, y_i, z_i) = (a, b, b)$ then $\Delta(\bm{x}, \bm{y}, \bm{z}) = \{i\}$ and consequently $V_1^{\overline{M}} = \{i\}$, $\Delta^M(\bm{x}, \bm{y}, \bm{z}) = \emptyset$.*

(b) *If $(x_i, y_i, z_i) = (b, a, b)$ then $\Delta(\bm{x}, \bm{y}, \bm{z}) = \{i\}$ and consequently $V_2^{\overline{M}} = \{i\}$, $\Delta^M(\bm{x}, \bm{y}, \bm{z}) = \emptyset$.*

(c) *If $(x_i, y_i, z_i) = (b, b, a)$ then $V_3^{\overline{M}} = \{i\}$, $|\{x_j, y_j, z_j\}| \le 2$ for all $j \in V$ and $\Delta^M(\bm{x}, \bm{y}, \bm{z}) = \emptyset$.*

*Proof.*

**Part (a)** Suppose that $(x_i, y_i, z_i) = (a, b, b)$ and $\Delta(\bm{x}, \bm{y}, \bm{z})$ is a strict superset of $\{i\}$. Let us define $\bm{u} = \text{Mn}_3(\bm{x}, \bm{y}, \bm{z})$. It can be checked that $\text{Mj}_1(\bm{x}, \bm{x}, \bm{u}) = \text{Mj}_2(\bm{x}, \bm{x}, \bm{u}) = \bm{x}$ and $\text{Mn}_3(\bm{x}, \bm{x}, \bm{u}) = \bm{u}$. Therefore, if we define $\bm{x}' = \bm{x}$ and $\bm{u}' = \bm{u}$ then the following identities will hold:

$$
\begin{aligned}
\text{Mj}_1(\bm{x}', \bm{y}, \bm{z}) &= \text{Mj}_1(\bm{x}, \bm{y}, \bm{z}) & \text{Mj}_1(\bm{x}, \bm{x}', \bm{u}') &= \bm{x}' \\
\text{Mj}_2(\bm{x}', \bm{y}, \bm{z}) &= \text{Mj}_2(\bm{x}, \bm{y}, \bm{z}) & \text{Mj}_2(\bm{x}, \bm{x}', \bm{u}') &= \bm{x}' \\
\text{Mn}_3(\bm{x}', \bm{y}, \bm{z}) &= \bm{u}' & \text{Mn}_3(\bm{x}, \bm{x}', \bm{u}') &= \text{Mn}_3(\bm{x}, \bm{y}, \bm{z})
\end{aligned}
$$

Let us modify $\bm{x}'$ and $\bm{u}'$ by setting $x'_i = u'_i = b$. It can be checked that the identities above still hold. By Lemma 25, $\bm{x}' \in \text{dom} f$. We also have $\delta(\bm{x}', \bm{y}, \bm{z}) < \delta(\bm{x}, \bm{y}, \bm{z})$, so Assumption 4 gives

$$f(\text{Mj}_1(\bm{x}, \bm{y}, \bm{z})) + f(\text{Mj}_2(\bm{x}, \bm{y}, \bm{z})) + f(\bm{u}') \le f(\bm{x}') + f(\bm{y}) + f(\bm{z}) \tag{22}$$

This implies, in particular, that $\bm{u}' \in \text{dom} f$. We have $(\bm{x}, \bm{x}', \bm{u}') \prec (\bm{x}, \bm{y}, \bm{z})$ since $\Delta(\bm{x}, \bm{x}', \bm{u}') = \{i\}$ and we assumed that $\Delta(\bm{x}, \bm{y}, \bm{z})$ is a strict superset of $\{i\}$. Therefore, Assumption 4 gives

$$f(\bm{x}') + f(\bm{x}') + f(\text{Mn}_3(\bm{x}, \bm{y}, \bm{z})) \le f(\bm{x}) + f(\bm{x}') + f(\bm{u}') \tag{23}$$

Summing (22) and (23) gives (18).

**Part (b)** Suppose that $(x_i, y_i, z_i) = (b, a, b)$ and $\Delta(\bm{x}, \bm{y}, \bm{z})$ is a strict subset of $V - \{i\}$. Let $\bm{u} = \text{Mn}_3(\bm{x}, \bm{y}, \bm{z})$. If we define $\bm{y}' = \bm{y}$ and $\bm{u}' = \bm{u}$ then the following identities will hold:

$$
\begin{aligned}
\text{Mj}_1(\bm{x}, \bm{y}', \bm{z}) &= \text{Mj}_1(\bm{x}, \bm{y}, \bm{z}) & \text{Mj}_1(\bm{y}, \bm{y}', \bm{u}') &= \bm{y}' \\
\text{Mj}_2(\bm{x}, \bm{y}', \bm{z}) &= \text{Mj}_2(\bm{x}, \bm{y}, \bm{z}) & \text{Mj}_2(\bm{y}, \bm{y}', \bm{u}') &= \bm{y}' \\
\text{Mn}_3(\bm{x}, \bm{y}', \bm{z}) &= \bm{u}' & \text{Mn}_3(\bm{y}, \bm{y}', \bm{u}') &= \text{Mn}_3(\bm{x}, \bm{y}, \bm{z})
\end{aligned}
$$

Let us modify $\bm{y}'$ and $\bm{u}'$ by setting $y'_i = u'_i = b$. It can be checked that the identities above still hold. The rest of the proof is analogous to the proof for part (a).

**Part (c)** Suppose that $(x_i, y_i, z_i) = (b, b, a)$ and (c) does not hold. Let $\bm{u} = \text{Mn}_3(\bm{x}, \bm{y}, \bm{z})$. If we define $\bm{z}' = \bm{z}$ and $\bm{u}' = \bm{u}$ then the following identities will hold:

$$
\begin{aligned}
\text{Mj}_1(\bm{x}, \bm{y}, \bm{z}') &= \text{Mj}_1(\bm{x}, \bm{y}, \bm{z}) & \text{Mj}_1(\bm{z}, \bm{z}', \bm{u}') &= \bm{z}' \\
\text{Mj}_2(\bm{x}, \bm{y}, \bm{z}') &= \text{Mj}_2(\bm{x}, \bm{y}, \bm{z}) & \text{Mj}_2(\bm{z}, \bm{z}', \bm{u}') &= \bm{z}' \\
\text{Mn}_3(\bm{x}, \bm{y}, \bm{z}') &= \bm{u}' & \text{Mn}_3(\bm{z}, \bm{z}', \bm{u}') &= \text{Mn}_3(\bm{x}, \bm{y}, \bm{z})
\end{aligned}
$$

Let us modify $\bm{z}'$ and $\bm{u}'$ by setting $z'_i = u'_i = b$. It can be checked that the identities above still hold.

We claim that $(*)$ $(\bm{z}, \bm{z}', \bm{u}') \prec (\bm{x}, \bm{y}, \bm{z})$. Indeed, since (c) does not hold we must have one of the following:

- $V_3^{\overline{M}}$ contains another node $j$ besides $i$. Then $(*)$ holds since $|\{z_j, z'_j, u'_j\}| = 1 < |\{x_j, y_j, z_j\}| = 2$.

- $|\{x_j, y_j, z_j\}| = 3$ for some $j \in V$. Then $(*)$ holds since $|\{z_j, z'_j, u'_j\}| \leq 2$.

- $|\Delta^M(\boldsymbol{x}, \boldsymbol{y}, \boldsymbol{z})| \geq 1$. Then $(*)$ holds since $|\Delta(\boldsymbol{z}, \boldsymbol{z}', \boldsymbol{u}')| = 1 \leq |\Delta^M(\boldsymbol{x}, \boldsymbol{y}, \boldsymbol{z})| \leq |\Delta(\boldsymbol{x}, \boldsymbol{y}, \boldsymbol{z})|$ and $|\Delta^M(\boldsymbol{z}, \boldsymbol{z}', \boldsymbol{u}')| = 0$.

The rest of the proof is analogous to the proof for part (a).

□

Next, we show that if $V^{\overline{M}}$ is non-empty then $V^M$ is empty. By Proposition 26 we know that in this case $\Delta^M(\boldsymbol{x}, \boldsymbol{y}, \boldsymbol{z})$ is empty. Thus, if $V^{\overline{M}} \neq \varnothing$ and $i \in V^M$ then we must have $(x_i, y_i, z_i) = (b, b, a)$. This case is eliminated by the following proposition.

**Proposition 27.** *For node $i \in V$ the following situations are impossible:*

S1 $(x_i, y_i, z_i) = (b, b, a)$, $(a, b) \in M$, $a \sqcup b = b$.

S2 $(x_i, y_i, z_i) = (b, b, a)$, $(a, b) \in M$, $a \sqcap b = b$.

*Proof.*
**Case S1** Let us define $\boldsymbol{u} = \mathtt{Mn}_3(\boldsymbol{x}, \boldsymbol{y}, \boldsymbol{z})$. By inspecting each case (17a)-(17d) and using equations (20) one can check that $\boldsymbol{u} \sqcup \boldsymbol{z} = \boldsymbol{z}$ and consequently $\boldsymbol{u} \sqcap \boldsymbol{z} = \boldsymbol{u}$. Therefore, if we define $\boldsymbol{z}' = \boldsymbol{z}$ and $\boldsymbol{u}' = \boldsymbol{u}$ then the following identities will hold:

$$\begin{array}{rclrcl}
\mathtt{Mj}_1(\boldsymbol{x}, \boldsymbol{y}, \boldsymbol{z}') & = & \mathtt{Mj}_1(\boldsymbol{x}, \boldsymbol{y}, \boldsymbol{z}) & \boldsymbol{u}' \sqcap \boldsymbol{z} & = & \mathtt{Mn}_3(\boldsymbol{x}, \boldsymbol{y}, \boldsymbol{z}) \\
\mathtt{Mj}_2(\boldsymbol{x}, \boldsymbol{y}, \boldsymbol{z}') & = & \mathtt{Mj}_2(\boldsymbol{x}, \boldsymbol{y}, \boldsymbol{z}) & \boldsymbol{u}' \sqcup \boldsymbol{z} & = & \boldsymbol{z}' \\
\mathtt{Mn}_3(\boldsymbol{x}, \boldsymbol{y}, \boldsymbol{z}') & = & \boldsymbol{u}'
\end{array}$$

Let us modify $\boldsymbol{z}'$ and $\boldsymbol{u}'$ by setting $z'_i = u'_i = b$, so that we have

$$\begin{array}{rl}
- \ a = z_i = u_i & = \mathtt{Mn}_3(x_i, y_i, z_i) \\
- \ b = z'_i = u'_i & = \mathtt{Mj}_{1,2}(x_i, y_i, z_i) \quad (= x_i = y_i)
\end{array} \qquad (a \sqcup b = b)$$

It can be checked that the identities above still hold. We have $\delta(\boldsymbol{x}, \boldsymbol{y}, \boldsymbol{z}') < \delta(\boldsymbol{x}, \boldsymbol{y}, \boldsymbol{z})$, so Assumption 4 gives

$$f(\mathtt{Mj}_1(\boldsymbol{x}, \boldsymbol{y}, \boldsymbol{z})) + f(\mathtt{Mj}_2(\boldsymbol{x}, \boldsymbol{y}, \boldsymbol{z})) + f(\boldsymbol{u}') \leq f(\boldsymbol{x}) + f(\boldsymbol{y}) + f(\boldsymbol{z}') \qquad (24)$$

assuming that $\boldsymbol{z}' \in \mathrm{dom}\, f$, and the fact that $\langle \sqcap, \sqcup \rangle$ is a multimorphism of $f$ gives

$$f(\mathtt{Mn}_3(\boldsymbol{x}, \boldsymbol{y}, \boldsymbol{z})) + f(\boldsymbol{z}') \leq f(\boldsymbol{u}') + f(\boldsymbol{z}) \qquad (25)$$

assuming that $\boldsymbol{u}' \in \mathrm{dom}\, f$. If $\boldsymbol{z}' \in \mathrm{dom}\, f$ then (24) implies that $\boldsymbol{u}' \in \mathrm{dom}\, f$; summing (24) and (25) gives (18). We thus assume that $\boldsymbol{z}' \notin \mathrm{dom}\, f$, then (25) implies that $\boldsymbol{u}' \notin \mathrm{dom}\, f$.

Let $C$ be a sufficiently large constant, namely $C > f(\boldsymbol{x}) + f(\boldsymbol{y}) + f(\boldsymbol{z})$. Consider function

$$g(\boldsymbol{u}) = \min_{d \in D}\{[d = a] \cdot C + f(d, \boldsymbol{u})\} \qquad \forall \boldsymbol{u} \in D^{\hat{V}} \qquad (26)$$

where $\hat{V} = V - \{i\}$, $[\cdot]$ is the Iverson bracket (it is 1 if its argument is true, and 0 otherwise) and we assumed for simplicity of notation that $i$ corresponds to the first argument of $f$. For an assignment $\boldsymbol{w} \in V$ we denote $\hat{\boldsymbol{w}}$ to be the restriction of $\boldsymbol{w}$ to $\hat{V}$. We can write

$$g(\hat{\boldsymbol{z}}) = f(\boldsymbol{z}) + C \qquad g(\hat{\boldsymbol{x}}) = f(\boldsymbol{x}) \qquad g(\hat{\boldsymbol{y}}) = f(\boldsymbol{y}) \qquad g(\hat{\boldsymbol{u}}) = f(\boldsymbol{u}) + C$$

where the first equation holds since $(b, \hat{\boldsymbol{z}}) = \boldsymbol{z}' \notin \text{dom}\, f$ and the last equation holds since $(b, \hat{\boldsymbol{u}}) = \boldsymbol{u}' \notin \text{dom}\, f$. Assumption 4 gives

$$g(\text{Mj}_1(\hat{\boldsymbol{x}}, \hat{\boldsymbol{y}}, \hat{\boldsymbol{z}})) + g(\text{Mj}_2(\hat{\boldsymbol{x}}, \hat{\boldsymbol{y}}, \hat{\boldsymbol{z}})) + g(\text{Mn}_3(\hat{\boldsymbol{x}}, \hat{\boldsymbol{y}}, \hat{\boldsymbol{z}})) \leq g(\hat{\boldsymbol{x}}) + g(\hat{\boldsymbol{y}}) + g(\hat{\boldsymbol{z}})$$
$$g(\text{Mj}_1(\hat{\boldsymbol{x}}, \hat{\boldsymbol{y}}, \hat{\boldsymbol{z}})) + g(\text{Mn}_3(\hat{\boldsymbol{x}}, \hat{\boldsymbol{y}}, \hat{\boldsymbol{z}})) + [(f(\boldsymbol{u}) + C] \leq f(\boldsymbol{x}) + f(\boldsymbol{y}) + [f(\boldsymbol{z}) + C]$$

Therefore, $g(\text{Mj}_1(\hat{\boldsymbol{x}}, \hat{\boldsymbol{y}}, \hat{\boldsymbol{z}})) < C$, and thus $g(\text{Mj}_1(\hat{\boldsymbol{x}}, \hat{\boldsymbol{y}}, \hat{\boldsymbol{z}})) = f(b, \text{Mj}_1(\hat{\boldsymbol{x}}, \hat{\boldsymbol{y}}, \hat{\boldsymbol{z}})) = f(\text{Mj}_1(\boldsymbol{x}, \boldsymbol{y}, \boldsymbol{z}))$. Similarly, $g(\text{Mj}_2(\hat{\boldsymbol{x}}, \hat{\boldsymbol{y}}, \hat{\boldsymbol{z}})) = f(b, \text{Mj}_2(\hat{\boldsymbol{x}}, \hat{\boldsymbol{y}}, \hat{\boldsymbol{z}})) = f(\text{Mj}_2(\boldsymbol{x}, \boldsymbol{y}, \boldsymbol{z}))$, and hence the inequality above is equivalent to (18).

**Case S2** Let us define $\boldsymbol{u} = \text{Mn}_3(\boldsymbol{x}, \boldsymbol{y}, \boldsymbol{z})$. It can be checked that $\boldsymbol{z} \sqcap \boldsymbol{u} = \boldsymbol{z}$ and consequently $\boldsymbol{z} \sqcup \boldsymbol{u} = \boldsymbol{u}$. Therefore, if we define $\boldsymbol{z}' = \boldsymbol{z}$ and $\boldsymbol{u}' = \boldsymbol{u}$ then the following identities will hold:

$$\text{Mj}_1(\boldsymbol{x}, \boldsymbol{y}, \boldsymbol{z}') = \text{Mj}_1(\boldsymbol{x}, \boldsymbol{y}, \boldsymbol{z}) \qquad \boldsymbol{z} \sqcap \boldsymbol{u}' = \boldsymbol{z}'$$
$$\text{Mj}_2(\boldsymbol{x}, \boldsymbol{y}, \boldsymbol{z}') = \text{Mj}_2(\boldsymbol{x}, \boldsymbol{y}, \boldsymbol{z}) \qquad \boldsymbol{z} \sqcup \boldsymbol{u}' = \text{Mn}_3(\boldsymbol{x}, \boldsymbol{y}, \boldsymbol{z})$$
$$\text{Mn}_3(\boldsymbol{x}, \boldsymbol{y}, \boldsymbol{z}') = \boldsymbol{u}'$$

Let us modify $\boldsymbol{z}'$ and $\boldsymbol{u}'$ by setting $z'_i = u'_i = b$, so that we have

- $a = z_i = u_i = \text{Mn}_3(x_i, y_i, z_i)$
- $b = z'_i = u'_i = \text{Mj}_{1,2}(x_i, y_i, z_i) \quad (= x_i = y_i)$ $\hfill (a \sqcap b = b)$

It can be checked that the identities above still hold. The rest of the proof proceeds analogously to the proof for the case S1. $\square$

We are now ready to prove the following fact.

**Proposition 28.** *Set $V^{\overline{M}}$ is empty.*

*Proof.* Suppose that $V^{\overline{M}} \neq \varnothing$. As we just showed, we must have $V^M = \varnothing$. For each $i \in V$ we also have $|\{x_i, y_i, z_i\}| \neq 1$ by Proposition 23 and $|\{x_i, y_i, z_i\}| \neq 3$ by Proposition 26. Therefore, $V = V^{\overline{M}}$. Proposition 26 implies that each of the sets $V_1^{\overline{M}}$, $V_2^{\overline{M}}$, $V_3^{\overline{M}}$ contains at most one node, and furthermore $|V_1^{\overline{M}} \cup V_2^{\overline{M}}| \leq 1$. Since $|V| \geq 2$ by Proposition 23, we conclude that $V = \{i, j\}$ where $i \in V_3^{\overline{M}}$ and $j \in V_1^{\overline{M}} \cup V_2^{\overline{M}}$.

Suppose that $j \in V_1^{\overline{M}}$, then we have $\boldsymbol{x} = (b, a')$, $\boldsymbol{y} = (b, b')$, $\boldsymbol{z} = (a, b')$ where $\{a, b\}, \{a', b'\} \in \overline{M}$. Inequality (18) reduces to

$$f(b, b') + f(b, b') + f(a, a') \leq f(b, a') + f(b, b') + f(a, b') \tag{27}$$

We must have $f(a, a') + f(b, b') = f(a, b') + f(b, a')$, otherwise $(a, b)$ would have a soft incident edge in $G_\Gamma$ contradicting to Lemma 15(g). Therefore, (27) is an equality. The case $j \in V_2^{\overline{M}}$ is completely analogous. Proposition 28 is proved. $\square$

## 6.4 Eliminating cases (17b) and (17c)

Propositions 24 and 28 show that there must exist node $i \in V$ with $\mu(\{x_i, y_i, z_i\}) = \{x_i\}$ or $\mu(\{x_i, y_i, z_i\}) = \{y_i\}$. In this section we show that this leads to a contradiction, thus proving Theorem 22.

Consider variable $i \in V$ with $\mu(\{x_i, y_i, z_i\}) = \{a\} \neq \varnothing$. We say that another variable $j \in V - \{i\}$ is a *control variable* for $i$ if $\{x_j, y_j, z_j\} = \{\alpha, \beta\} \in \overline{M}$ and for any labelling $u \in \text{dom} f$ the following is true: $u_i = a$ iff $u_j = \alpha$. This implies the following property:

**Proposition 29.** *Suppose that variable $i \in V$ with $\mu(\{x_i, y_i, z_i\}) = \{a\}$ has a control variable. Let $u$, $v$, $w$ be a permutation of $x, y, z$ such that $u_i = a$. Then*

- *Any labelling obtained from one of the labellings in $\{u, \text{Mn}_3(x, y, z)\}$ by changing the label of $i$ from $a$ to $v_i$ or $w_i$ does not belong to $\text{dom} f$.*

- *Any labelling obtained from one of the labellings in $\{v, w, \text{Mj}_1(x, y, z), \text{Mj}_2(x, y, z)\}$ by changing the label of $i$ from $\{v_i, w_i\}$ to $a$ does not belong to $\text{dom} f$.*

Let $(f, x, y, z)$ be a valid instance and $i \in V$ be a variable with $\mu(\{x_i, y_i, z_i\}) \neq \varnothing$. If $i$ does not have a control variable then we can define another valid instance $(\bar{f}, \bar{x}, \bar{y}, \bar{z})$ with the set of variables $\bar{V} = V \cup \{j\}$, $j \neq V$ as follows:
$$\bar{f}(u) = f(\hat{u}) + g(u_i, u_j) \qquad \forall u \in D^{\bar{V}}$$
where $g$ is a binary function taken from the definition of the set $\mu(\{x_i, y_i, z_i\})$ and $\hat{u}$ is the restriction of $u$ to $V$. Labellings $\bar{x}, \bar{y}, \bar{z}$ are obtained by extending $x, y, z$ to $\bar{V}$ in the unique way so that $(\bar{f}, \bar{x}, \bar{y}, \bar{z})$ is a valid instance. Clearly, in the new instance variable $i$ does have a control variable. Furthermore, this transformation does not affect the satisfiability of the instance, and $\delta(x, y, z)$ is increased by 2. Such transformation will be used below; after introducing control variable $j$ we will "minimise out" variable $x_i$, which will decrease $\delta(x, y, z)$ by 3.

If $\mu(\{a, b, c\}) = \{c\}$ then we will illustrate this fact using the following diagram:

**Proposition 30.** *For node $i \in V$ the following situations are impossible:*

- **T1** $\mu(\{x_i, y_i, z_i\}) = \{y_i\}$, $(x_i, z_i) \in M$, $x_i \sqcap z_i = z_i$.

- **T2** $\mu(\{x_i, y_i, z_i\}) = \{y_i\}$, $(x_i, z_i) \in M$, $x_i \sqcup z_i = z_i$.

- **T3** $\mu(\{x_i, y_i, z_i\}) = \{x_i\}$, $(y_i, z_i) \in M$, $y_i \sqcup z_i = z_i$.

- **T4** $\mu(\{x_i, y_i, z_i\}) = \{x_i\}$, $(y_i, z_i) \in M$, $y_i \sqcap z_i = z_i$.

*Proof.* We will analyse cases T1-T4 separately, and will derive a contradiction in each case.

**Case T1** Let us define $u = \text{Mj}_2(x, y, z)$. It can be checked that $x \sqcap u = x$ and consequently $x \sqcup u = u$. Therefore, if we define $x' = x$ and $u' = u$ then the following identities will hold:

$$\begin{aligned} \text{Mj}_1(x', y, z) &= \text{Mj}_1(x, y, z) & x \sqcap u' &= x' \\ \text{Mj}_2(x', y, z) &= u' & x \sqcup u' &= \text{Mj}_2(x, y, z) = u \qquad (28) \\ \text{Mn}_3(x', y, z) &= \text{Mn}_3(x, y, z) & & \end{aligned}$$

Let us modify $\boldsymbol{x}', \boldsymbol{u}'$ by setting $x'_i = u'_i = \mathtt{Mj}_1(x_i, y_i, z_i)$ so that we have

$$\begin{array}{rll} a = x_i = u_i & = \mathtt{Mj}_2(x_i, y_i, z_i) & \\ b = x'_i = u'_i & = \mathtt{Mj}_1(x_i, y_i, z_i) \ (= z_i) & (a \sqcap b = b) \\ c & = \mathtt{Mn}_3(x_i, y_i, z_i) \ (= y_i) & \end{array}$$

where we denoted $(a, b, c) = (x_i, z_i, y_i)$. It can be checked that identities (28) still hold, and furthermore $\delta(\boldsymbol{x}', \boldsymbol{y}, \boldsymbol{z}) < \delta(\boldsymbol{x}, \boldsymbol{y}, \boldsymbol{z})$. Assumption 4 gives

$$f(\mathtt{Mj}_1(\boldsymbol{x}, \boldsymbol{y}, \boldsymbol{z})) + f(\boldsymbol{u}') + f(\mathtt{Mn}_3(\boldsymbol{x}, \boldsymbol{y}, \boldsymbol{z})) \ \leq \ f(\boldsymbol{x}') + f(\boldsymbol{y}) + f(\boldsymbol{z}) \tag{29}$$

assuming that $\boldsymbol{x}' \in \mathrm{dom}\, f$, and the fact that $\langle \sqcap, \sqcup \rangle$ is a multimorphism of $f$ gives

$$f(\boldsymbol{x}') + f(\mathtt{Mj}_2(\boldsymbol{x}, \boldsymbol{y}, \boldsymbol{z})) \ \leq \ f(\boldsymbol{x}) + f(\boldsymbol{u}') \tag{30}$$

assuming that $\boldsymbol{u}' \in \mathrm{dom}\, f$. If $\boldsymbol{x}' \in \mathrm{dom}\, f$ then (29) implies that $\boldsymbol{u}' \in \mathrm{dom}\, f$; summing (29) and (30) gives (18). We thus assume that $\boldsymbol{x}' \notin \mathrm{dom}\, f$, then (30) implies that $\boldsymbol{u}' \notin \mathrm{dom}\, f$.

Let us add a control variable for $i$ using the transformation described above. For simplicity, we do not change the notation, so we assume that $V$ now contains a control variable for $i$ and $\boldsymbol{x}, \boldsymbol{y}, \boldsymbol{z}, \boldsymbol{u}, \boldsymbol{x}', \boldsymbol{u}'$ have been extended to the new set accordingly. We have $\delta(\boldsymbol{x}, \boldsymbol{y}, \boldsymbol{z}) = \delta_{\min} + 2$.

Let $C$ be a sufficiently large constant, namely $C > f(\boldsymbol{x}) + f(\boldsymbol{y}) + f(\boldsymbol{z})$. Consider function

$$g(\boldsymbol{u}) = \min_{d \in D}\{[d = a] \cdot C + f(d, \boldsymbol{u})\} \qquad \forall \boldsymbol{u} \in D^{\hat{V}} \tag{31}$$

where $\hat{V} = V - \{i\}$, $[\cdot]$ is the Iverson bracket (it returns 1 if its argument is true and 0 otherwise) and we assumed for simplicity of notation that $i$ corresponds to the first argument of $f$. For an assignment $\boldsymbol{w} \in V$ we denote $\hat{\boldsymbol{w}}$ to be the restriction of $\boldsymbol{w}$ to $\hat{V}$. We can write

$$g(\hat{\boldsymbol{x}}) = f(\boldsymbol{x}) + C \qquad g(\hat{\boldsymbol{y}}) = f(\boldsymbol{y}) \qquad g(\hat{\boldsymbol{z}}) = f(\boldsymbol{z}) \qquad g(\hat{\boldsymbol{u}}) = f(\boldsymbol{u}) + C \tag{32}$$

To show the first equation, observe that the minimum in (31) cannot be achieved at $d = b$ since $(b, \hat{\boldsymbol{x}}) = \boldsymbol{x}' \notin \mathrm{dom}\, f$, and also the minimum cannot be achieved at $d = c$ by Proposition 29. Therefore, $g(\hat{\boldsymbol{x}}) = g(a, \hat{\boldsymbol{x}}) = f(\boldsymbol{x}) + C$. Other equations can be derived similarly.

Clearly, $(g, \hat{\boldsymbol{x}}, \hat{\boldsymbol{y}}, \hat{\boldsymbol{z}})$ is a valid instance and $\delta(\hat{\boldsymbol{x}}, \hat{\boldsymbol{y}}, \hat{\boldsymbol{z}}) = \delta_{\min} - 1$, so Assumption 4 gives

$$\begin{aligned} g(\mathtt{Mj}_1(\hat{\boldsymbol{x}}, \hat{\boldsymbol{y}}, \hat{\boldsymbol{z}})) + g(\mathtt{Mj}_2(\hat{\boldsymbol{x}}, \hat{\boldsymbol{y}}, \hat{\boldsymbol{z}})) + g(\mathtt{Mn}_3(\hat{\boldsymbol{x}}, \hat{\boldsymbol{y}}, \hat{\boldsymbol{z}})) & \leq g(\hat{\boldsymbol{x}}) + g(\hat{\boldsymbol{y}}) + g(\hat{\boldsymbol{z}}) \\ g(\mathtt{Mj}_1(\hat{\boldsymbol{x}}, \hat{\boldsymbol{y}}, \hat{\boldsymbol{z}})) + [f(\boldsymbol{u}) + C] + g(\mathtt{Mn}_3(\hat{\boldsymbol{x}}, \hat{\boldsymbol{y}}, \hat{\boldsymbol{z}})) & \leq [f(\boldsymbol{x}) + C] + f(\boldsymbol{y}) + f(\boldsymbol{z}) \end{aligned}$$

Therefore, $g(\mathtt{Mj}_1(\hat{\boldsymbol{x}}, \hat{\boldsymbol{y}}, \hat{\boldsymbol{z}})) < C$, and thus $g(\mathtt{Mj}_1(\hat{\boldsymbol{x}}, \hat{\boldsymbol{y}}, \hat{\boldsymbol{z}})) = f(b, \mathtt{Mj}_1(\hat{\boldsymbol{x}}, \hat{\boldsymbol{y}}, \hat{\boldsymbol{z}})) = f(\mathtt{Mj}_1(\boldsymbol{x}, \boldsymbol{y}, \boldsymbol{z}))$. (Note, labelling $(c, \mathtt{Mj}_1(\hat{\boldsymbol{x}}, \hat{\boldsymbol{y}}, \hat{\boldsymbol{z}}))$ is not in $\mathrm{dom}\, f$ by Proposition 29.) Similarly, $g(\mathtt{Mn}_3(\hat{\boldsymbol{x}}, \hat{\boldsymbol{y}}, \hat{\boldsymbol{z}})) = f(c, \mathtt{Mn}_3(\hat{\boldsymbol{x}}, \hat{\boldsymbol{y}}, \hat{\boldsymbol{z}})) = f(\mathtt{Mn}_3(\boldsymbol{x}, \boldsymbol{y}, \boldsymbol{z}))$, and hence the inequality above is equivalent to (18).

**Case** T2 Let us define $\boldsymbol{u} = \mathtt{Mj}_1(\boldsymbol{x}, \boldsymbol{y}, \boldsymbol{z})$. It can be checked that $\boldsymbol{u} \sqcup \boldsymbol{x} = \boldsymbol{x}$ and consequently $\boldsymbol{u} \sqcap \boldsymbol{x} = \boldsymbol{u}$. Therefore, if we define $\boldsymbol{x}' = \boldsymbol{x}$ and $\boldsymbol{u}' = \boldsymbol{u}$ then the following identities will hold:

$$\begin{aligned} \mathtt{Mj}_1(\boldsymbol{x}', \boldsymbol{y}, \boldsymbol{z}) & = \boldsymbol{u}' & \boldsymbol{u}' \sqcap \boldsymbol{x} & = \mathtt{Mj}_1(\boldsymbol{x}, \boldsymbol{y}, \boldsymbol{z}) = \boldsymbol{u} \\ \mathtt{Mj}_2(\boldsymbol{x}', \boldsymbol{y}, \boldsymbol{z}) & = \mathtt{Mj}_2(\boldsymbol{x}, \boldsymbol{y}, \boldsymbol{z}) & \boldsymbol{u}' \sqcup \boldsymbol{x} & = \boldsymbol{x}' \\ \mathtt{Mn}_3(\boldsymbol{x}', \boldsymbol{y}, \boldsymbol{z}) & = \mathtt{Mn}_3(\boldsymbol{x}, \boldsymbol{y}, \boldsymbol{z}) & & \end{aligned}$$

Let us modify $x', u'$ by setting $x'_i = u'_i = \mathtt{Mj}_2(x_i, y_i, z_i)$ so that we have

$$\begin{cases} a = x_i = u_i &= \mathtt{Mj}_1(x_i, y_i, z_i) \\ b = x'_i = u'_i &= \mathtt{Mj}_2(x_i, y_i, z_i) \ (= z_i) \\ c &= \mathtt{Mn}_3(x_i, y_i, z_i) \ (= y_i) \end{cases} \qquad (a \sqcup b = b)$$

It can be checked that the identities above still hold. The rest of the proof proceeds analogously to the proof for the case T1.

**Case T3** Let us define $u = \mathtt{Mj}_1(x, y, z)$. It can be checked that $u \sqcup y = y$ and consequently $u \sqcap y = u$. Therefore, if we define $y' = y$ and $u' = u$ then the following identities will hold:

$$\begin{aligned}
\mathtt{Mj}_1(x, y', z) &= u' & u' \sqcap y &= \mathtt{Mj}_1(x, y, z) = u \\
\mathtt{Mj}_2(x, y', z) &= \mathtt{Mj}_2(x, y, z) & u' \sqcup y &= y' \\
\mathtt{Mn}_3(x, y', z) &= \mathtt{Mn}_3(x, y, z)
\end{aligned}$$

Let us modify $y', u'$ by setting $y'_i = u'_i = \mathtt{Mj}_2(x_i, y_i, z_i)$ so that we have

$$\begin{cases} a = y_i = u_i &= \mathtt{Mj}_1(x_i, y_i, z_i) \\ b = y'_i = u'_i &= \mathtt{Mj}_2(x_i, y_i, z_i) \ (= z_i) \\ c &= \mathtt{Mn}_3(x_i, y_i, z_i) \ (= x_i) \end{cases} \qquad (a \sqcup b = b)$$

It can be checked that the identities above still hold. The rest of the proof proceeds analogously to the proof for the case T1.

**Case T4** Let us define $u = \mathtt{Mj}_2(x, y, z)$. It can be checked that $y \sqcap u = y$ and consequently $y \sqcup u = u$. Therefore, if we define $y' = y$ and $u' = u$ then the following identities will hold:

$$\begin{aligned}
\mathtt{Mj}_1(x, y', z) &= \mathtt{Mj}_1(x, y, z) & y \sqcap u' &= y' \\
\mathtt{Mj}_2(x, y', z) &= u' & y \sqcup u' &= \mathtt{Mj}_2(x, y, z) = u \\
\mathtt{Mn}_3(x, y', z) &= \mathtt{Mn}_3(x, y, z)
\end{aligned}$$

Let us modify $y', u'$ by setting $y'_i = u'_i = \mathtt{Mj}_1(x_i, y_i, z_i)$ so that we have

$$\begin{cases} a = y_i = u_i &= \mathtt{Mj}_2(x_i, y_i, z_i) \\ b = y'_i = u'_i &= \mathtt{Mj}_1(x_i, y_i, z_i) \ (= z_i) \\ c &= \mathtt{Mn}_3(x_i, y_i, z_i) \ (= x_i) \end{cases} \qquad (a \sqcap b = b)$$

It can be checked that the identities above still hold. The rest of the proof proceeds analogously to the proof for the case T1. $\square$

There are two possible cases remaining: $\mu(\{x_i, y_i, z_i\}) = \{y_i\}$, $\{x_i, z_i\} \in \overline{M}$ or $\mu(\{x_i, y_i, z_i\}) = \{x_i\}$, $\{y_i, z_i\} \in \overline{M}$. They are eliminated by the next two propositions; we use a slightly different argument.

**Proposition 31.** *For node $i \in V$ the following situation is impossible:*

T5 $\mu(\{x_i, y_i, z_i\}) = \{y_i\}$, $\{x_i, z_i\} \in \overline{M}$.

*Proof.* For a labelling $w \in D^V$ let $\hat{w}$ be the restriction of $w$ to $V - \{i\}$. Two cases are possible.

**Case 1** $(\text{Mj}_2(\hat{x}, \hat{y}, \hat{z}), \hat{y}, \hat{z}) \prec (\hat{x}, \hat{y}, \hat{z})$. Let us define $u = \text{Mj}_2(x, y, z)$ and $v = \text{Mj}_2(u, y, z)$. It can be checked that $\text{MJN}(u, v, z) = (u, v, z)$. [3] Therefore, if we define $z' = z$ and $u' = u$ then the following identities will hold:

$$\text{Mj}_1(x, y, z') = \text{Mj}_1(x, y, z) \qquad \text{Mj}_1(u', v, z) = \text{Mj}_2(x, y, z) = u \qquad v = \text{Mj}_2(u', y, z)$$
$$\text{Mj}_2(x, y, z') = u' \qquad \text{Mj}_2(u', v, z) = v$$
$$\text{Mn}_3(x, y, z') = \text{Mn}_3(x, y, z) \qquad \text{Mn}_3(u', v, z) = z'$$

Let us modify $z'$ and $u'$ according to the following diagram:

$$\begin{array}{l} a = z_i = u_i = \text{Mj}_2(x_i, y_i, z_i) \ (= v_i) \\ b = z'_i = u'_i = \text{Mj}_1(x_i, y_i, z_i) \ (= x_i) \\ c = \phantom{z'_i = u'_i} = \text{Mn}_3(x_i, y_i, z_i) \ (= y_i) \end{array}$$

It can be checked that the identities above still hold. The assumption of Case 1 gives $(u', y, z) \prec (x, y, z)$ (note that $u'_i = x_i$). Therefore, the fact that $v = \text{Mj}_2(u', y, z)$ and Assumption 4 give the following relationship: $(*)$ if $u' \in \text{dom} f$ then $v \in \text{dom} f$.

We have $\delta(x, y, z') < \delta(x, y, z)$ and $\delta(u', v, z) < \delta(x, y, z)$, so Assumption 4 gives

$$f(\text{Mj}_1(x, y, z)) + f(u') + f(\text{Mn}_3(x, y, z)) \leq f(x) + f(y) + f(z') \tag{33}$$

assuming that $z' \in \text{dom} f$, and

$$f(\text{Mj}_2(x, y, z)) + f(v) + f(z') \leq f(u') + f(v) + f(z) \tag{34}$$

assuming that $u', v \in \text{dom} f$. If $z' \in \text{dom} f$ then (33) implies that $u' \in \text{dom} f$, and so $(*)$ implies that $v \in \text{dom} f$. Summing (33) and (34) gives (18). We thus assume that $z' \notin \text{dom} f$, then we have $u' \notin \text{dom} f$. (If $u' \in \text{dom} f$ then $(*)$ gives $v \in \text{dom} f$, and equation (34) then gives $z' \in \text{dom} f$ - a contradiction.)

The rest of the argument proceeds similar to that for the case T1. Let us add a control variable for $i$ (again, without changing the notation). Consider function

$$g(u) = \min_{d \in D}\{[d = a] \cdot C + f(d, u)\} \qquad \forall u \in D^{\hat{V}}$$

where $\hat{V} = V - \{i\}$ and $C > f(x) + f(y) + f(z)$ is a sufficiently large constant. We can write

$$g(\hat{z}) = f(z) + C \qquad g(\hat{x}) = f(x) \qquad g(\hat{y}) = f(y) \qquad g(\hat{u}) = f(u) + C$$

Clearly, $(g, \hat{x}, \hat{y}, \hat{z})$ is a valid instance and $\delta(\hat{x}, \hat{y}, \hat{z}) = \delta_{\min} - 1$, so Assumption 4 gives

$$g(\text{Mj}_1(\hat{x}, \hat{y}, \hat{z})) + g(\text{Mj}_2(\hat{x}, \hat{y}, \hat{z})) + g(\text{Mn}_3(\hat{x}, \hat{y}, \hat{z})) \leq g(\hat{x}) + g(\hat{y}) + g(\hat{z})$$
$$g(\text{Mj}_1(\hat{x}, \hat{y}, \hat{z})) + [f(u) + C] + g(\text{Mn}_3(\hat{x}, \hat{y}, \hat{z})) \leq f(x) + f(y) + [f(z) + C]$$

Therefore, $g(\text{Mj}_1(\hat{x}, \hat{y}, \hat{z})) < C$, and thus $g(\text{Mj}_1(\hat{x}, \hat{y}, \hat{z})) = f(b, \text{Mj}_1(\hat{x}, \hat{y}, \hat{z})) = f(\text{Mj}_1(x, y, z))$. Similarly, $g(\text{Mn}_3(\hat{x}, \hat{y}, \hat{z})) = f(c, \text{Mn}_3(\hat{x}, \hat{y}, \hat{z})) = f(\text{Mn}_3(x, y, z))$, and hence the inequality above is equivalent to (18).

**Case 2** $(\text{Mj}_2(\hat{x}, \hat{y}, \hat{z}), \hat{y}, \hat{z}) \not\prec (\hat{x}, \hat{y}, \hat{z})$. This implies, in particular, the following condition:

---

[3] If $u_j = v_j$ then obviously $\text{MJN}(u_j, v_j, z_j) = (u_j, v_j, z_j)$; suppose that $u_j \neq v_j$. This implies $u_j \neq x_j$ and $u_j \neq y_j$ (if $u_j = y_j$ then we would have $v_j = \text{Mj}_2(u_j, u_j, z_j) = u_j$). Therefore, $u_j = z_j$. We must have $v_j = \text{Mj}_2(z_j, y_j, z_j) = y_j$ since $v_j \neq u_j = z_j$. Thus, $\text{MJN}(u_j, v_j, z_j) = \text{MJN}(z_j, y_j, z_j) = (\alpha, y_j, \beta)$. We have $\{\{z_j, y_j, z_j\}\} = \{\{\alpha, y_j, \beta\}\}$, and so $\alpha = \beta = z_j$.

(∗) if $|\{x_j, y_j, z_j\}| = 3$ for $j \in V - \{i\}$ then $\mathtt{Mj}_2(x_j, y_j, z_j) = x_j$.

It is easy to check that $\Delta(\mathtt{Mj}_2(\hat{\boldsymbol{x}}, \hat{\boldsymbol{y}}, \hat{\boldsymbol{z}}), \hat{\boldsymbol{y}}, \hat{\boldsymbol{z}}) \subseteq \Delta(\hat{\boldsymbol{x}}, \hat{\boldsymbol{y}}, \hat{\boldsymbol{z}})$. Indeed, consider node $j \in V - \{i\}$ with $\mathtt{Mj}_2(x_j, y_j, z_j) \neq y_j$; we need to show that $x_j \neq y_j$. If $|\{x_j, y_j, z_j\}| = 3$ then this follows from (∗), so it remains to consider the case when $\mathtt{MJN}(x_j, y_j, z_j)$ is defined via (17d) (case (17a) was eliminated by Proposition 28). We then have $\mathtt{Mj}_2(x_j, y_j, z_j) = x_j \sqcup y_j$, and so $x_j \sqcup y_j \neq y_j$ clearly implies $x_j \neq y_j$.

We thus must have $\Delta(\mathtt{Mj}_2(\hat{\boldsymbol{x}}, \hat{\boldsymbol{y}}, \hat{\boldsymbol{z}}), \hat{\boldsymbol{y}}, \hat{\boldsymbol{z}}) = \Delta(\hat{\boldsymbol{x}}, \hat{\boldsymbol{y}}, \hat{\boldsymbol{z}})$, otherwise the assumption of Case 2 would not hold. This implies the following:

(∗∗) if $x_j \neq y_j$ for $j \in V - \{i\}$ then $\mathtt{Mj}_2(x_j, y_j, z_j) \neq y_j$.

Let us define $\boldsymbol{u} = \mathtt{Mj}_1(\boldsymbol{x}, \boldsymbol{y}, \boldsymbol{z})$, and let $\boldsymbol{x}', \boldsymbol{u}'$ be the labellings obtained from $\boldsymbol{x}, \boldsymbol{u}$ by setting $x'_i = u'_i = z_i$, so that we have

$$\begin{array}{l} a = x_i = u_i \; = \mathtt{Mj}_1(x_i, y_i, z_i) \\ b = x'_i = u'_i \; = \mathtt{Mj}_2(x_i, y_i, z_i) \; (= z_i) \\ c \qquad\qquad\quad = \mathtt{Mn}_3(x_i, y_i, z_i) \; (= y_i) \end{array}$$

We claim that the following identities hold:

$$\begin{array}{rclcrcl} \mathtt{Mj}_1(\boldsymbol{x}', \boldsymbol{y}, \boldsymbol{z}) & = & \boldsymbol{u}' & \quad & \boldsymbol{x} \sqcap \boldsymbol{u}' & = & \mathtt{Mj}_1(\boldsymbol{x}, \boldsymbol{y}, \boldsymbol{z}) = \boldsymbol{u} \\ \mathtt{Mj}_2(\boldsymbol{x}', \boldsymbol{y}, \boldsymbol{z}) & = & \mathtt{Mj}_2(\boldsymbol{x}, \boldsymbol{y}, \boldsymbol{z}) & \quad & \boldsymbol{x} \sqcup \boldsymbol{u}' & = & \boldsymbol{x}' \\ \mathtt{Mn}_3(\boldsymbol{x}', \boldsymbol{y}, \boldsymbol{z}) & = & \mathtt{Mn}_3(\boldsymbol{x}, \boldsymbol{y}, \boldsymbol{z}) & & & & \end{array}$$

Indeed, we need to show that $x_j \sqcup u_j = x_j$ for $j \in V - \{i\}$. If $\mathtt{MJN}(x_j, y_j, z_j)$ was defined via (17b) then $\mathtt{Mj}_2(x_j, y_j, z_j) = y_j \sqcup z_j \neq x_j$ contradicting to condition (∗). Similarly, if it was defined via (17c) then $\mathtt{Mj}_2(x_j, y_j, z_j) = x_j \sqcup z_j = z_j \neq x_j$ again contradicting to condition (∗). (Note, in the latter case $x_j \sqcup z_j = z_j$ since by Proposition 30 we cannot have $\{x_j, z_j\} \in M$.) We showed that $\mathtt{MJN}(x_j, y_j, z_j)$ must be determined via (17d), so $u_j = \mathtt{Mj}_1(x_j, y_j, z_j) = x_j \sqcap y_j$ and $\mathtt{Mj}_2(x_j, y_j, z_j) = x_j \sqcup y_j$. If $x_j = y_j$ then the claim $x_j \sqcup u_j = x_j$ is trivial. If $x_j \neq y_j$ then condition (∗∗) implies $x_j \sqcup y_j \neq y_j$, and consequently $x_j \sqcup y_j = x_j$, $u_j = x_j \sqcap y_j = y_j$ and $x_j \sqcup u_j = x_j \sqcup y_j = x_j$, as claimed.

The rest of the proof proceeds analogously to the proof for the case T1. □

**Proposition 32.** *For node $i \in V$ the following situation is impossible:*

T6 $\mu(\{x_i, y_i, z_i\}) = \{x_i\}$, $\{y_i, z_i\} \in \overline{M}$.

*Proof.* Let us define $\boldsymbol{u} = \mathtt{Mj}_2(\boldsymbol{x}, \boldsymbol{y}, \boldsymbol{z})$ and $\boldsymbol{v} = \mathtt{Mj}_2(\boldsymbol{u}, \boldsymbol{x}, \boldsymbol{z})$. It can be checked that $\mathtt{MJN}(\boldsymbol{v}, \boldsymbol{u}, \boldsymbol{z}) = (\boldsymbol{v}, \boldsymbol{u}, \boldsymbol{z})$. [4] Therefore, if we define $\boldsymbol{z}' = \boldsymbol{z}$ and $\boldsymbol{u}' = \boldsymbol{u}$ then the following identities will hold:

$$\begin{array}{rclcrclcrcl} \mathtt{Mj}_1(\boldsymbol{x}, \boldsymbol{y}, \boldsymbol{z}') & = & \mathtt{Mj}_1(\boldsymbol{x}, \boldsymbol{y}, \boldsymbol{z}) & \quad & \mathtt{Mj}_1(\boldsymbol{v}, \boldsymbol{u}', \boldsymbol{z}) & = & \boldsymbol{v} & \quad & \boldsymbol{v} & = & \mathtt{Mj}_2(\boldsymbol{u}', \boldsymbol{x}, \boldsymbol{z}) \\ \mathtt{Mj}_2(\boldsymbol{x}, \boldsymbol{y}, \boldsymbol{z}') & = & \boldsymbol{u}' & \quad & \mathtt{Mj}_2(\boldsymbol{v}, \boldsymbol{u}', \boldsymbol{z}) & = & \mathtt{Mj}_2(\boldsymbol{x}, \boldsymbol{y}, \boldsymbol{z}) = \boldsymbol{u} & & & & \\ \mathtt{Mn}_3(\boldsymbol{x}, \boldsymbol{y}, \boldsymbol{z}') & = & \mathtt{Mn}_3(\boldsymbol{x}, \boldsymbol{y}, \boldsymbol{z}) & \quad & \mathtt{Mn}_3(\boldsymbol{v}, \boldsymbol{u}', \boldsymbol{z}) & = & \boldsymbol{z}' & & & & \end{array}$$

---

[4] If $u_j = v_j$ then obviously $\mathtt{MJN}(v_j, u_j, z_j) = (v_j, u_j, z_j)$; suppose that $u_j \neq v_j$. This implies $u_j \neq x_j$ (otherwise we would have $v_j = \mathtt{Mj}_2(u_j, u_j, z_j) = u_j$). If $\mathtt{MJN}(x_j, y_j, z_j)$ is determined via (17b) then $\{y_j, z_j\} \in \overline{M}$ by Proposition 30 and so $u_j = z_j$ and $v_j = z_j$. It remains to consider the case when it is determined via (17d) (cases (17a) and (17c) have been eliminated). We have $u_j = x_j \sqcup y_j = y_j$ since $u_j \neq x_j$, and so $v_j = \mathtt{Mj}_2(y_j, x_j, z_j) = y_j \sqcup x_j = x_j$ since $v_j \neq u_j = y_j$ (clearly, $\mathtt{Mj}_2(y_j, x_j, z_j)$ is also determined via (17d)). We thus have $\mathtt{MJN}(v_j, u_j, z_j) = \mathtt{MJN}(x_j, y_j, z_j) = (\alpha, u_j, z_j)$. Condition $\{\{v_j, u_j, z_j\}\} = \{\{\alpha, u_j, z_j\}\}$ implies that $\alpha = v_j$.

Let us modify $z'$ and $u'$ according to the following diagram:

$$\begin{array}{l} a = z_i = u_i \phantom{'} = \mathtt{Mj}_2(x_i, y_i, z_i) \; (= v_i) \\ b = z'_i = u'_i = \mathtt{Mj}_1(x_i, y_i, z_i) \; (= y_i) \\ c \phantom{= z'_i = u'_i} = \mathtt{Mn}_3(x_i, y_i, z_i) \; (= x_i) \end{array}$$

It can be checked that the identities above still hold. It suffices to show that $(u', x, z) \prec (x, y, z)$, then the proof will be analogous to the proof for the Case 1 of T5.

Consider node $j \in V - \{i\}$. We will show next that $j$ satisfies the following:

(a) If $j \in \Delta(u', x, z)$ then $j \in \Delta(x, y, z)$. In other words, if $u'_j \ne x_j$ then $y_j \ne x_j$.

(b) If $j \in \Delta^M(u', x, z)$ then $j \in \Delta^M(x, y, z)$. Namely, if $(u'_j, x_j, z_j) = (a, b, b)$ or $(u'_j, x_j, z_j) = (b, a, b)$ where $\{a, b\} \in M$ then $u'_i = y_i$ and thus $(x_i, y_i, z_i) = (b, a, b)$ or $(x_i, y_i, z_i) = (a, b, b)$ respectively.

(c) $\mu(\{u'_j, x_j, z_j\}) \ne \{u'_j\}$.

This will imply the claim since $(u'_i, x_i, z_i) = (y_i, x_i, z_i) \prec (x_i, y_i, z_i)$ due to the fourth component in (19).

If $\mathtt{MJN}(x_j, y_j, z_j)$ is determined via (17b) then we must have $\{y_j, z_j\} \in \overline{M}$ by Proposition 30, and so $u'_j = \mathtt{Mj}_2(x_j, y_j, z_j) = z_j$. Checking (a-c) is then straightforward.

It remains to consider the case when $\mathtt{MJN}(x_j, y_j, z_j)$ is determined via (17d) - all other cases have been eliminated. Condition (c) then clearly holds, and $u'_j = \mathtt{Mj}_2(x_j, y_j, z_j) = x_j \sqcup y_j$. If $u'_j = x_j$ then (a,b) are trivial since their preconditions do not hold. It is also straightforward to check that (a,b) hold if $u'_j = y_j \ne x_j$. $\square$

## 7 Proof of Theorem 11

In this section we present an algorithm for minimising instances from $\mathsf{VCSP}(\Gamma)$. The idea for the algorithm and some of the proof techniques have been influenced by the techniques used by Takhanov [46] for proving the absence of *arithmetical deadlocks* in certain instances. However, the algorithm itself is very different from Takhanov's approach. (The latter does not rely on submodular minimization algorithms; instead, it performs a reduction to an optimization problem in a perfect graph).

Let $f : \mathcal{D} \to \overline{\mathbb{Q}}_+$ be the function to be minimised, $V$ be the set of its variables (which we will also call nodes), and $D_i$ be the domain of variable $i \in V$ with $\mathcal{D} = \times_{i \in V} D_i$. In the beginning all domains are the same ($D_i = D$), but as the algorithm progresses we will allow $D_i$ to become different for different $i \in V$. As a consequence, operations $\sqcap, \sqcup$ may act differently on different components of vectors $x, y \in \mathcal{D}$. We denote $\sqcap_i, \sqcup_i : D_i \times D_i \to D_i$ to be the $i$-th operations of $\langle \sqcap, \sqcup \rangle$. Similarly, we denote by $\mathtt{Mj}_{1i}, \mathtt{Mj}_{2i}, \mathtt{Mn}_{3i} : D_i \times D_i \times D_i \to D_i$ to be the $i$-th operations of $\langle \mathtt{Mj}_1, \mathtt{Mj}_2, \mathtt{Mn}_3 \rangle$.

We denote by $P$ the collection of sets $P = (P_i)_{i \in V}$ where $P_i = \{\{a, b\} \mid a, b \in D_i, a \ne b\}$. We denote by $M$ a collection of subsets $M = (M_i)_{i \in V}$, $M_i \subseteq P_i$, and $\overline{M} = (\overline{M_i})_{i \in V}$, $\overline{M_i} = P_i - M_i$. We now extend Definition 8 as follows.

**Definition 33.** *Let $\langle \sqcap, \sqcup \rangle$ and $\langle \mathtt{Mj}_1, \mathtt{Mj}_2, \mathtt{Mn}_3 \rangle$ be collections of binary and ternary operations respectively.*

- *Pair $\langle \sqcap, \sqcup \rangle$ is an STP on $M$ if for all $i \in V$ pair $\langle \sqcap_i, \sqcup_i \rangle$ is an STP on $M_i$.*

- *Triple $\langle \mathtt{Mj}_1, \mathtt{Mj}_2, \mathtt{Mn}_3 \rangle$ is an MJN on $\overline{M}$ if for all $i \in V$ triple $\langle \mathtt{Mj}_{1i}, \mathtt{Mj}_{2i}, \mathtt{Mn}_{3i} \rangle$ is an MJN on $\overline{M_i}$.*

We will assume without loss of generality that $\langle \sqcap_i, \sqcup_i \rangle$ is non-commutative on any $\{a, b\} \in \overline{M}_i$ (if not, we can simply add such $\{a, b\}$ to $M_i$).

We are now ready to present the algorithm; it will consist of three stages.

### Stage 1: Decomposition into binary relations

Since the instance admits a majority polymorphism (see Section 7.1), every cost function $f$ can be decomposed [1] into unary relations $\rho_i \subseteq D_i$, $i \in D_i$ and binary relations $\rho_{ij} \subseteq D_i \times D_j$, $i,j \in V$, $i \neq j$ such that
$$\boldsymbol{x} \in \text{dom}\, f \quad\Leftrightarrow\quad [x_i \in \rho_i \,\forall i \in V] \quad\text{and}\quad [(x_i,x_j) \in \rho_{ij} \,\forall i,j \in V, i \neq j]$$
We will always assume that binary relations are symmetric, i.e. $(x,y) \in \rho_{ij} \Leftrightarrow (y,x) \in \rho_{ji}$. We use the following notation for relations:

- If $\rho_{ij} \in D_i \times D_j$, $X \subseteq D_i$ and $Y \subseteq D_j$ then
$$\rho_{ij}(X,\cdot) = \{y \mid \exists x \in X \text{ s.t. } (x,y) \in \rho_{ij}\} \qquad \rho_{ij}(\cdot,Y) = \{x \mid \exists y \in Y \text{ s.t. } (x,y) \in \rho_{ij}\}$$
If $X = \{x\}$ and $Y = \{y\}$ then these two sets will be denoted as $\rho_{ij}(x,\cdot)$ and $\rho_{ij}(\cdot,y)$ respectively.

- If $\rho \in D_1 \times D_2$ and $\rho' \in D_2 \times D_3$ then we define their composition as
$$\rho \circ \rho' = \{(x,z) \in D_1 \times D_3 \mid \exists y \in D_2 \text{ s.t. } (x,y) \in \rho, (y,z) \in \rho'\}$$

In the first stage we establish *strong 3-consistency* using the standard constraint-processing techniques [16] so that the resulting relations satisfy

| | | | |
|---|---|---|---|
| (arc-consistency) | $\{x \mid (\exists y)(x,y) \in \rho_{ij}\}$ | $= \rho_i$ | $\forall$ distinct $i,j \in V$ |
| (path-consistency) | $\rho_{ik}(x,\cdot) \cap \rho_{jk}(y,\cdot)$ | $\neq \varnothing$ | $\forall$ distinct $i,j,k \in V$, $(x,y) \in \rho_{ij}$ |

It is known that in the presence of a majority polymorphism strong 3-consistency is equivalent to global consistency [31]; that is $\text{dom}\, f$ is empty iff all $\rho_i$ and $\rho_{ij}$ are empty. Using this fact, it is not difficult to show that the strong 3-consistency relations $\rho_i, \rho_{ij}$ are uniquely determined by $f$ via
$$\rho_i = \{x_i \mid \boldsymbol{x} \in \text{dom}\, f\} \qquad \rho_{ij} = \{(x_i,x_j) \mid \boldsymbol{x} \in \text{dom}\, f\}$$
The second equation implies that any polymorphism of $f$ is also a polymorphism of $\rho_{ij}$.

From now on we will assume that $D_i = \rho_i$ for all $i \in V$. This can be achieved by reducing sets $D_i$ if necessary. We will also assume that all sets $D_i$ are non-empty.

### Stage 2: Modifying $M$ and $\langle \sqcap, \sqcup \rangle$

The second stage of the algorithm works by iteratively growing sets $M_i$ and simultaneously modifying operations $\langle \sqcap_i, \sqcup_i \rangle$ so that (i) $\langle \sqcap_i, \sqcup_i \rangle$ is still a conservative pair which is commutative on $M_i$ and non-commutative on $\overline{M}_i$, and (ii) $\langle \sqcap, \sqcup \rangle$ is a multimorphism of $f$. It stops when we get $M_i = P_i$ for all $i \in V$.

We now describe one iteration. First, we identify subset $U \subseteq V$ and subsets $A_i, B_i \subseteq D_i$ for each $i \in U$ using the following algorithm:

1: pick node $k \in V$ and pair $\{a,b\} \in \overline{M}_k$. (If they do not exist, terminate and go to Stage 3.)
2: set $U = \{k\}$, $A_k = \{a\}$, $B_k = \{b\}$
3: **while** there exists $i \in V - U$ such that $\rho_{ki}(A_k, \cdot) \cap \rho_{ki}(B_k, \cdot) = \varnothing$ **do**
4:    add $i$ to $U$, set $A_i = \rho_{ki}(A_k, \cdot)$, $B_i = \rho_{ki}(B_k, \cdot)$
   // compute "closure" of sets $A_i$ for $i \in U$
5:    **while** there exists $a \in D_k - A_k$ s.t. $a \in \rho_{ki}(\cdot, A_i)$ for some $i \in U - \{k\}$ **do**
6:      add $a$ to $A_k$, set $A_j = \rho_{kj}(A_k, \cdot)$ for all $j \in U - \{k\}$
7:    **end while**
   // compute "closure" of sets $B_i$ for $i \in U$
8:    **while** there exists $b \in D_k - B_k$ s.t. $b \in \rho_{ki}(\cdot, B_i)$ for some $i \in U - \{k\}$ **do**
9:      add $b$ to $B_k$, set $B_j = \rho_{kj}(B_k, \cdot)$ for all $j \in U - \{k\}$
10:    **end while**
   // done
11: **end while**
12: **return** set $U \subseteq V$ and sets $A_i, B_i \subseteq D_i$ for $i \in U$

**Lemma 34.** *Sets $U$ and $A_i, B_i$ for $i \in U$ produced by the algorithm have the following properties:*

*(a) Sets $A_i$ and $B_i$ for $i \in U$ are disjoint.*

*(b) $\{a,b\} \in \overline{M}_i$ for all $i \in U$, $a \in A_i$, $b \in B_i$.*

*(c) $\rho_{ki}(A_k, \cdot) = A_i$, $\rho_{ki}(B_k, \cdot) = B_i$, $\rho_{ki}(\cdot, A_i) = A_k$, $\rho_{ki}(\cdot, B_i) = B_k$ for all $i \in U - \{k\}$ where $k$ is the node chosen in line 1.*

*(d) Suppose that $i \in U$ and $j \in \overline{U} \equiv V - U$. If $(c,x) \in \rho_{ij}$ where $c \in A_i \cup B_i$ and $x \in D_j$ then $(d,x) \in \rho_{ij}$ for all $d \in A_i \cup B_i$.*

To complete the iteration, we modify sets $M_i$ and operations $\sqcap_i, \sqcup_i$ for each $i \in U$ as follows:

- add all pairs $\{a,b\}$ to $M_i$ where $a \in A_i$, $b \in B_i$.

- redefine $a \sqcap_i b = b \sqcap_i a = a$, $a \sqcup_i b = b \sqcup_i a = b$ for all $a \in A_i, b \in B_i$

**Lemma 35.** *The new pair of operations $\langle \sqcap, \sqcup \rangle$ is a multimorphism of $f$.*

A proof of Lemmas 34 and 35 is given in the next section. They imply that all steps are well-defined, and upon termination the algorithm produces a pair $\langle \sqcap, \sqcup \rangle$ which is an STP multimorphism of $f$.

### Stage 3: Reduction to a submodular minimisation problem

At this stage we have an STP multimorphism. Hence, the instance can be solved by Theorem 5.

### 7.1 Algorithm's correctness

First, we show that $f$ admits a majority polymorphism $\mu$ using the argument from [46]. Define

$$\bar{\mu}(\boldsymbol{x},\boldsymbol{y},\boldsymbol{z}) = [(\boldsymbol{y} \sqcup \boldsymbol{x}) \sqcap (\boldsymbol{y} \sqcup \boldsymbol{z})] \sqcap (\boldsymbol{x} \sqcup \boldsymbol{z})$$
$$\mu(\boldsymbol{x},\boldsymbol{y},\boldsymbol{z}) = \mathtt{Mj}_1(\bar{\mu}(\boldsymbol{x},\boldsymbol{y},\boldsymbol{z}), \bar{\mu}(\boldsymbol{y},\boldsymbol{z},\boldsymbol{x}), \bar{\mu}(\boldsymbol{z},\boldsymbol{x},\boldsymbol{y}))$$

Suppose that $\{x,y,z\} = \{a,b\} \in P_i$. It can be checked that $\bar{\mu}_i(x,y,z)$ acts as the majority operation if $\{a,b\} \in M_i$, and $\bar{\mu}_i(x,y,z) = x$ if $\{a,b\} \in \overline{M}_i$. This implies that $\mu_i$ acts as the majority operation on $P_i$.

**Proposition 36.** *If $\{a, b\} \in \overline{M}_i$, $\{a', b'\} \in P_j$ and $(a, a'), (b, b') \in \rho_{ij}$, where $i, j$ are distinct nodes in $V$, then exactly one of the following holds:*

(i) $(a, b'), (b, a') \in \rho_{ij}$

(ii) $(a, b'), (b, a') \notin \rho_{ij}$ and $\{a', b'\} \in \overline{M}_j$

*Proof.* First, suppose that $\{a', b'\} \in M_j$. We need to show that case (i) holds. Operations $\sqcap_i, \sqcup_i$ are non-commutative on $\{a, b\}$, while $\sqcap_j, \sqcup_j$ are commutative on $\{a', b'\}$. It is easy to check that

$$\{(a,b) \sqcap (a',b'), (a',b') \sqcap (a,b), (a,b) \sqcup (a',b'), (a',b') \sqcup (a,b)\} = \{(a,a'), (a,b'), (a',b), (a',b')\}$$

Since $\sqcap, \sqcup$ are polymorphisms of $\rho_{ij}$, all assignments involved in the equation above belong to $\rho_{ij}$. Thus, (i) holds.

Now suppose $\{a', b'\} \in \overline{M}_j$. We then have

$$\text{Mn}_3((a,a'), (b,b'), (a,b')) = (b,a') \qquad \text{Mn}_3((a,a'), (b,b'), (b,a')) = (a,b')$$

$\text{Mn}_3$ is a polymorphism of $\rho_{ij}$, therefore if one of the assignments $(a, b'), (b, a')$ belongs to $\rho_{ij}$ then the other one also belongs to $\rho_{ij}$. This proves the proposition. $\square$

### 7.1.1 Proof of Lemma 34(a-c)

It follows from construction that during all stages of the algorithm there holds

$$\rho_{ki}(A_k, \cdot) = A_i, \quad \rho_{ki}(B_k, \cdot) = B_i \qquad \forall i \in U - \{k\} \tag{35}$$

Strong 3-consistency also implies that sets $A_i, B_i$ for $i \in U$ are non-empty. Clearly, properties (a) and (b) of Lemma 34 hold after initialization (line 2). Let us prove that each step of the algorithm preserves these two properties. Note, property (a) together with (35) imply that $(a, b') \notin \rho_{ki}$ if $a \in A_k$, $b' \in B_i$, and $(b, a') \notin \rho_{ki}$ if $b \in B_k$, $a' \in A_i$, where $i \in U - \{k\}$.

First, consider line 4, i.e. adding $i$ to $U$ with $A_i = \rho_{ki}(A_k, \cdot)$, $B_i = \rho_{ki}(B_k, \cdot)$. Property (a) for node $i$ follows from the precondition of line 3; let us show (b) for node $i$. Suppose that $a' \in A_i$, $b' \in B_i$, then there exist $a \in A_k$, $b \in B_k$ such that $(a, a'), (b, b') \in \rho_{ki}$. We have $(a, b') \notin \rho_{ki}$, so by Proposition 36 we get $\{a', b'\} \in \overline{M}$.

Now consider line 6, i.e. adding $a$ to $A_k$ and updating $A_j$ for $j \in U - \{k\}$ accordingly. We denote $A_j^\circ$ and $A_j$ to be respectively the old and the new set for node $j \in U$. There must exist node $i \in U - \{k\}$ and element $a' \in A_i^\circ$ such that $(a, a') \in \rho_{ki}$. We prove below that properties (a) and (b) are preserved for nodes $k, i$ and all nodes $j \in U - \{k, i\}$.

**Node $k$** It is clear that $a \notin B_k$, otherwise we would have $a' \in \rho_{ki}(B_k, \cdot) = B_i$ contradicting to condition $A_i^\circ \cap B_i = \varnothing$. Thus, property (a) for node $k$ holds. Consider element $b \in B_k$. By arc-consistency there exists element $b' \in \rho_{ki}(b, \cdot) \subseteq B_i$. From property (b) we get $\{a', b'\} \in \overline{M}_i$. We also have $(b, a') \notin \rho_{ki}$ since $A_i^\circ \cap \rho_{ki}(B_k, \cdot) = A_i^\circ \cap B_i = \varnothing$. By Proposition 36 we get $\{a, b\} \in \overline{M}_k$. Thus, property (b) holds for node $k$.

**Node $i$** Let us prove that $A_i \cap B_i = \varnothing$. Suppose not, then $(a, b') \in \rho_{ki}$ for some $b' \in B_i$. There must exist $b \in B_k$ with $(b, b') \in \rho_{ki}$. We have $\rho_{ki} \cap (\{a,b\} \times \{a',b'\}) = \{(a,a'), (b,b'), (a,b')\}$ and $\{a',b'\} \in \overline{M}_i$, which is a contradiction by Proposition 36. This proves property (a) for node $i$.

Property (b) for node $i$ follows from property (a) for nodes $k, i$, property (b) for node $k$, and Proposition 36.

**Node $j \in U - \{k, i\}$** Let us prove that $A_j \cap B_j = \varnothing$. Suppose not, then $(a, y) \in \rho_{kj}$ for some $y \in B_j$. There must exist $b \in B_k$ with $(b, y) \in \rho_{kj}$, and $b' \in B_i$ with $(b, b') \in \rho_{ki}$. We also have $a' \in A_i^\circ =$ $\rho_{ki}(A_k^\circ, \cdot)$, therefore there must exist $c \in A_k^\circ$ with $(c, a') \in \rho_{ki}$, and $x \in A_{kj}^\circ$ with $(c, x) \in \rho_{kj}$. It can be seen that

$$\rho_{ki} \cap (\{a, c, b\} \times \{a', b'\}) = \{(a, a'), (c, a'), (b, b')\} \qquad \rho_{kj} \cap (\{a, c, b\} \times \{x, y\}) = \{(a, y), (c, x), (b, y)\}$$

Indeed, all listed assignments belong to $\rho_{ki}$ or $\rho_{kj}$ by construction; we need to show that remaining assignments do not belong to these relations. We have $(a, b'), (c, b'), (b, a') \notin \rho_{ki}$ since we have already established property (a) for nodes $k$ and $i$. We also have $(c, y), (b, x) \notin \rho_{kj}$ since $A_k^\circ \cap B_k = \varnothing$ and $A_j^\circ \cap B_j = \varnothing$. Combining it with the fact that $\{x, y\} \in \overline{M}$ and using Proposition 36 gives that $(a, x) \notin \rho_{kj}$.

Consider relation $\beta_{ij} = \rho'_{ik} \circ \rho_{kj}$ where $\rho'_{ik} = \{(d', d) \in \rho_{ik} \mid d \in \{a, b, c\}\}$. It is easy to check that $(a', x), (a', y), (b', y) \in \beta_{ij}$ and $(b', x) \notin \beta_{ij}$. We have $\{a', b'\} \in \overline{M}_i$ and $\{x, y\} \in \overline{M}_j$, so $\text{Mn}_3((a', x), (a', y), (b', y)) = (b', x)$. Clearly, $\text{Mn}_3$ is a polymorphism of $\rho'_{ik}$ and $\beta_{ij}$, therefore we must have $(b', x) \in \beta_{ij}$ - a contradiction. This proves property (a) for node $j$.

Property (b) for node $j$ follows from property (a) for nodes $k, j$, property (b) for node $k$, and Proposition 36.

**Concluding remark** We showed that throughout the algorithm sets $U, A_i, B_i$ satisfy properties (a,b) and equation (35). It is easy to see that after running lines 5-7 we also have $\rho_{ki}(\cdot, A_i) = A_k$, and after running lines 8-10 we have $\rho_{ki}(\cdot, B_i) = B_k$. Thus, property (c) holds upon termination, which concludes the proof of Lemma 34(a-c).

### 7.1.2 Proof of Lemma 34(d)

First, we will prove the following claim:

**Proposition 37.** *Suppose that $(a, x), (b, x), (c, y) \in \rho_{ij}$ where $i \in U$, $j \in \overline{U}$, $a \in A_i$, $b \in B_i$, $c \in A_i \cup B_i$, $x, y \in D_j$. Then $(a, y), (b, y), (c, x) \in \rho_{ij}$.*

*Proof.* We claim that there exists a relation $\gamma_i \subseteq D_i \times D_i$ with the following properties:

(i) $\gamma_i$ is an equivalence relation, i.e. there exists a unique partitioning $\pi[\gamma_i] = \{C_1, \ldots, C_p\}$ of $D_i$ such that $(x, y) \in \gamma_i$ for $x, y \in D_i$ iff $x$ and $y$ belong to the same partition of $\pi[\gamma_i]$;

(ii) $A_i \in \pi[\gamma_i]$ and $B_i \in \pi[\gamma_i]$;

(iii) operation $\text{Mn}_{3i}$ is a polymorphism of $\gamma_i$.

Indeed, for $i = k$ such relation can be constructed as follows. Let us set $\gamma_k = \{(a, a) \mid a \in D_k\}$ and iteratively update it via $\gamma_k := \gamma_k \circ \rho_{ki} \circ \rho_{ik}$ for $i \in U - \{k\}$. Set $\gamma_i$ will never shrink; we stop when no such operation can change $\gamma_k$. Clearly, at this point $\gamma_i$ is an equivalence relation. By comparing this scheme with lines 5-10 of the algorithm we conclude that (ii) holds. Finally, (iii) follows from the fact that polymorphisms are preserved under compositions. If $i \in U - \{k\}$ then we take $\gamma_i = \rho_{ik} \circ \gamma_k \rho_{ki}$; (i)-(iii) then follow from property (c) of Lemma 34.

We are now ready to prove Proposition 37. We can assume that $x \neq y$, otherwise the claim is trivial. Assume that $c \in A_i$ (the case $c \in B_i$ is analogous). Suppose that $(b, y) \notin \rho_{ij}$. We have $\{b, c\} \in \overline{M}$, so Proposition 36 implies that $\{x, y\} \in \overline{M}$. Consider relation $\gamma'_i = \{(x, y) \in \gamma_i \mid y \notin B_i - \{b\})\}$. Polymorphisms in property (iii) are conservative, therefore they are polymorphisms of $\gamma'_i$ as well. Define relation $\beta_{ij} = \gamma'_i \circ \rho_{ij} \subseteq D_i \times D_j$, then $\text{Mn}_3$ is a polymorphism of $\beta_{ij}$. It is easy to check that $(a, y), (a, x), (b, x) \in \beta_{ij}$. Operation $\text{Mn}_3$ is a polymorphism of $\beta_{ij}$ and it acts as the minority operation on $\{a, b\} \in \overline{M}$ and $\{x, y\} \in \overline{M}$, therefore $\text{Mn}_3((a, y), (a, x), (b, x)) = (b, y) \in \beta_{ij}$. This implies that $(b, y) \in \rho_{ij}$, contradicting to the assumption made earlier. We showed that we must have $(b, y) \in \rho_{ij}$. The fact that $\{a, b\} \in \overline{M}$ and Proposition 36 then imply that $(a, y) \in \rho_{ij}$. Finally, the fact that $\{c, b\} \in \overline{M}$ and Proposition 36 imply that $(c, x) \in \rho_{ij}$. Proposition 37 is proved. □

We can now prove Lemma 34(d) under the following assumption:

(∗) Sets $\rho_{ij}(A_i,\cdot)$ and $\rho_{ij}(B_i,\cdot)$ have non-empty intersection.

(This assumption clearly holds if $i = k$, otherwise the algorithm wouldn't have terminated; we will later show that (∗) holds for nodes $i \in U - \{k\}$ as well.)

First, let us prove that $\rho_{ij}(A_i,\cdot) = \rho_{ij}(B_i,\cdot)$. Suppose that $y \in \rho_{ij}(A_i,\cdot)$, then $(c,y) \in \rho_{ij}$ for some $c \in A_i$. From assumption (∗) we get that there exist $a \in A_i$, $b \in B_i$, $x \in D_j$ such that $(a,x), (b,x) \in \rho_{ij}$. Proposition 37 implies that $(b,y) \in \rho_{ij}$, and thus $\rho_{ij}(A_i,\cdot) \subseteq \rho_{ij}(B_i,\cdot)$. By symmetry we also have $\rho_{ij}(B_i,\cdot) \subseteq \rho_{ij}(A_i,\cdot)$, implying $\rho_{ij}(A_i,\cdot) = \rho_{ij}(B_i,\cdot)$.

Second, let us prove that if $(a,x) \in \rho_{ij}$ where $a \in A_i$, $x \in D_j$ then $(c,x) \in \rho_{ij}$ for all $c \in B_i$. (We call this claim [AB]). As we showed in the previous paragraph, there exists $b \in B_i$ such that $(b,x) \in \rho_{ij}$. We can also select $y \in D_j$ such that $(c,y) \in \rho_{ij}$. Proposition 37 implies that $(c,x) \in \rho_{ij}$, as desired.

A symmetrical argument shows that if $(b,x) \in \rho_{ij}$ where $b \in B_i$, $x \in D_j$ then $(c,x) \in \rho_{ij}$ for all $c \in A_i$ [BA]. By combining facts [AB] and [BA] we obtain that if $(a,x) \in \rho_{ij}$ where $a \in A_i$, $x \in D_j$ then $(c,x) \in \rho_{ij}$ for all $c \in A_i$ [AA], and also that if $(b,x) \in \rho_{ij}$ where $b \in B_i$, $x \in D_j$ then $(c,x) \in \rho_{ij}$ for all $c \in B_i$ [BB].

We have proved Lemma 34(d) assuming that (∗) holds (and in particular, for $i = k$). It remains to show that (∗) holds for $i \in U - \{k\}$. Let us select $(a', x) \in \rho_{ij}$ where $a' \in A_i$, $x, y \in D_j$. By strong 3-consistency there exists $a \in D_k$ such that $(a, a') \in \rho_{ki}$ and $(a, x) \in \rho_{kj}$. By Lemma 34(c) we get that $a \in A_k$. As we have just shown, there exists $b \in B_k$ such that $(b, x) \in \rho_{kj}$. By strong 3-consistency there exists $b' \in D_i$ such that $(b, b') \in \rho_{ki}$ and $(b', x) \in \rho_{ij}$. By Lemma 34(c) we get that $b' \in B_i$. We have shown that $x \in \rho_{ij}(A_i,\cdot)$ and $x \in \rho_{ij}(B_i,\cdot)$, which proves (∗).

### 7.1.3 Proof of Lemma 35

Suppose we have an arc- and path-consistent instance with an STP on $M$ and MJN on $\overline{M}$ and non-empty subset $U$ with $A_i, B_i \subseteq D_i$ for $i \in U$ that satisfy properties (a-d) of Lemma 34 (where node $k \in U$ is fixed). Let us denote $M^\circ$ and $M$ to be the set before and after the update respectively. Similarly, $\langle \sqcap^\circ, \sqcup^\circ \rangle$ and $\langle \sqcap, \sqcup \rangle$ denote operations before and after the update. We need to show that

$$f(\boldsymbol{x} \sqcap \boldsymbol{y}) + f(\boldsymbol{x} \sqcup \boldsymbol{y}) \le f(\boldsymbol{x}) + f(\boldsymbol{y}) \qquad \text{if } \boldsymbol{x}, \boldsymbol{y} \in \operatorname{dom} f \tag{36}$$

For a vector $\boldsymbol{z} \in \mathcal{D}$ and subset $S \subseteq V$ we denote $\boldsymbol{z}^S$ to be the restriction of $\boldsymbol{z}$ to $S$. Given $\boldsymbol{x}, \boldsymbol{y} \in \mathcal{D}$, denote

$$\delta(\boldsymbol{x}, \boldsymbol{y}) = \begin{cases} 0 & \text{if } \boldsymbol{x}^U \sqcap \boldsymbol{y}^U = \boldsymbol{x}^U \sqcap^\circ \boldsymbol{y}^U \\ 1 & \text{otherwise} \end{cases} \qquad \Delta(\boldsymbol{x}, \boldsymbol{y}) = \{i \in \overline{U} \mid x_i \ne y_i\}$$

Note, if $\delta(\boldsymbol{x}, \boldsymbol{y}) = 0$ then $\boldsymbol{x} \sqcap \boldsymbol{y} = \boldsymbol{x} \sqcap^\circ \boldsymbol{y}$ and $\boldsymbol{x} \sqcup \boldsymbol{y} = \boldsymbol{x} \sqcup^\circ \boldsymbol{y}$, so the claim is trivial. Let us introduce a partial order $\preceq$ on pairs $(\boldsymbol{x}, \boldsymbol{y})$ as the lexicographical order on vector $(|\Delta(\boldsymbol{x}, \boldsymbol{y})|, \delta(\boldsymbol{x}, \boldsymbol{y}))$ (the first component is more significant than the second). We use induction on this order. The base of the induction follows from the following lemma.

**Lemma 38.** *Condition* (36) *holds for all* $\boldsymbol{x}, \boldsymbol{y} \in \operatorname{dom} f$ *with* $|\Delta(\boldsymbol{x}, \boldsymbol{y})| \le 1$.

*Proof.* We can assume that $\delta(\boldsymbol{x}, \boldsymbol{y}) = 1$, otherwise the claim holds trivially. Thus, there exists node $i \in U$ such that either $x_i \in A_i, y_i \in B_i$ or $x_i \in B_i, y_i \in A_i$, Lemma 34(c) implies that either $x_i \in A_i, y_i \in B_i$ for all $i \in U$ or $x_i \in B_i, y_i \in A_i$ for all $i \in U$. Therefore, from the definition of operations $\sqcap, \sqcup$ we get $\{\boldsymbol{x}^U \sqcap \boldsymbol{y}^U, \boldsymbol{x}^U \sqcup \boldsymbol{y}^U\} = \{\boldsymbol{x}^U, \boldsymbol{y}^U\}$. Also, we have $\boldsymbol{x} \sqcap^\circ \boldsymbol{y}, \boldsymbol{x} \sqcup^\circ \boldsymbol{y} \in \operatorname{dom} f$, so Lemma 34(c) gives $\{\boldsymbol{x}^U \sqcap^\circ \boldsymbol{y}^U, \boldsymbol{x}^U \sqcup^\circ \boldsymbol{y}^U\} = \{\boldsymbol{x}^U, \boldsymbol{y}^U\}$.

If $|\Delta(x,y)| = 0$ then $\{x \sqcap y, x \sqcup y\} = \{x, y\}$ and so the claim holds trivially. Let us assume that $\Delta(x,y) = \{j\}$. We will write $x = (x^U, x_j, z)$ and $y = (y^U, y_j, z)$ where $z = x^{\overline{U}-\{j\}} = y^{\overline{U}-\{j\}}$. Denote $z^{01} = (x^U, y_j, z)$ and $z^{10} = (y^U, x_j, z)$. Clearly, we have either $\{x \sqcap y, x \sqcup y\} = \{x, y\}$ or $\{x \sqcap y, x \sqcup y\} = \{z^{01}, z^{10}\}$. We can assume that the latter condition holds, otherwise (36) is a trivial equality. By Lemma 34(d) we have $(x_i, y_j), (y_i, x_j) \in \rho_{ij}$ for all $i \in U$, therefore $z^{01}, x^{10} \in \text{dom} f$. Two cases are possible:

**Case 1** $\{x_j, y_j\} \in M_j$, so $\sqcap_j^\circ, \sqcup_j^\circ$ are commutative on $\{x_j, y_j\}$. Thus, we must have either $\{x \sqcap^\circ y, x \sqcup^\circ y\} = \{z^{01}, z^{10}\}$ or $\{y \sqcap^\circ x, y \sqcup^\circ x\} = \{z^{01}, z^{10}\}$. Using the fact that $\langle \sqcap^\circ, \sqcup^\circ \rangle$ is a multimorphism of $f$, we get in each case the desired inequality:

$$f(z^{01}) + f(z^{01}) \leq f(x) + f(y)$$

**Case 2** $\{x_j, y_j\} \in \overline{M}_j$. It can be checked that applying operations $\langle \text{Mj}_1, \text{Mj}_2, \text{Mn}_3 \rangle$ to $(x, y, z^{01})$ gives $(z^{01}, z^{01}, z^{10})$, therefore

$$f(z^{01}) + f(z^{01}) + f(z^{10}) \leq f(x) + f(y) + f(z^{01})$$

which is equivalent to (36). □

**Proposition 39.** *If $x, y \in \text{dom} f$ and $\delta(x, y) = 1$ then either $\delta(x \sqcup y, y) = 0$ or $\delta(x, x \sqcup y) = 0$.*

*Proof.* Using the same argumentation as in the proof of Lemma 38 we conclude that $\{x^U \sqcap y^U, x^U \sqcup y^U\} = \{x^U, y^U\}$. If $x^U \sqcup y^U = x^U$ then $\delta(x \sqcup y, y) = 0$, and if $x^U \sqcup y^U = y^U$ then $\delta(x, x \sqcup y) = 0$. □

We now proceed with the induction argument. Suppose that $\Delta(x, y) \geq 2$. We can assume without loss of generality that $\delta(x, y) = 1$, otherwise the claim is trivial. Denote

$$\begin{aligned} X &= \{i \in \Delta(x, y) \mid x_i \sqcap y_i = x_i, \ x_i \sqcup y_i = y_i\} \\ Y &= \{i \in \Delta(x, y) \mid x_i \sqcap y_i = y_i, \ x_i \sqcup y_i = x_i\} \end{aligned}$$

We have $|X \cup Y| \geq 2$, so by Proposition 39 at least one of the two cases below holds:

**Case 1** $|X| \geq 2$ or $|X| = 1, \delta(x \sqcup y, y) = 0$. It can be checked that $(x \sqcup y) \sqcap y = y$. Therefore, if we define $x' = x \sqcup y, y' = y$ then the following identities hold:

$$x \sqcap y' = x \sqcap y \qquad x \sqcup y' = x' \qquad x' \sqcap y = y' \qquad x' \sqcup y = x \sqcup y \qquad (37)$$

Let us select node $s \in X$ and modify $y'$ by setting $y'_s = x_s$. (Note that we have $x'_s = x_s$.) It can be checked that (37) still holds. We have

- $(x, y') \prec (x, y)$ since $\Delta(x, y') = \Delta(x, y) - \{s\}$, and
- $(x', y) \prec (x, y)$ since $\Delta(x', y) = \Delta(x, y) - (X - \{s\})$; if $X - \{s\}$ is empty then $\delta(x', y) < \delta(x, y)$.

Thus, by the induction hypothesis

$$f(x \sqcap y) + f(x') \leq f(x) + f(y') \qquad (38)$$

assuming that $y' \in \text{dom} f$, and

$$f(y') + f(x \sqcup y) \leq f(x') + f(y) \qquad (39)$$

assuming that $\boldsymbol{x}' \in \mathrm{dom}\,f$. If $\boldsymbol{y}' \in \mathrm{dom}\,f$ then Inequality (38) implies that $\boldsymbol{x}' \in \mathrm{dom}\,f$, and the claim then follows from summing (38) and (39). We now assume that $\boldsymbol{y}' \notin \mathrm{dom}\,f$; Inequality (39) then implies that $\boldsymbol{x}' \notin \mathrm{dom}\,f$.

Assume for simplicity of notation that $k$ corresponds to the first argument of $f$. Define instance $\hat{\mathcal{I}}$ with the set of nodes $\hat{V} = V - \{s\}$ and cost function

$$g(\boldsymbol{z}) = \min_{a \in D_s} \{u(a) + f(a, \boldsymbol{z})\} \qquad \forall \boldsymbol{z} \in \hat{\mathcal{D}} \equiv \bigotimes_{i \in \hat{V}} D_i$$

where $u(a)$ is the following unary cost function: $u(x_s) = 0$, $u(y_s) = C$ and $u(a) = \infty$ for $a \in D - \{x_s, y_s\}$. Here $C$ is a sufficiently large constant, namely $C > f(\boldsymbol{x}) + f(\boldsymbol{y})$. It is straightforward to check that unary relations $D_i, i \in \hat{V}$ and binary relations $\rho_{ij}, i, j \in \hat{V}, i \neq j$ are the unique arc- and path-consistent relations for $g$, i.e.

$$\rho_i = \{x_i \,|\, \boldsymbol{x} \in \mathrm{dom}\,g\} \quad \forall i \in \hat{V}, \qquad \rho_{ij} = \{(x_i, x_j) \,|\, \boldsymbol{x} \in \mathrm{dom}\,g\} \quad \forall i, j \in \hat{V}, i \neq j$$

This implies that set $U \subseteq \hat{V}$ and sets $A_i, B_i$ for $i \in U$ satisfy conditions (a-d) of Lemma 34 for instance $\hat{\mathcal{I}}$. Operations $\langle \sqcap^\circ, \sqcup^\circ \rangle$ and $\langle \mathrm{Mj}_1, \mathrm{Mj}_2, \mathrm{Mn}_3 \rangle$ are multimorphisms of functions $u$ (since they are conservative) and $f$ (by assumption), therefore they are also multimorphisms of $g$. Furthermore, if the modification in Stage 2 had been applied to instance $\hat{\mathcal{I}}$ and sets $U, A_i, B_i$ then it would give the same pair $\langle \sqcap, \sqcup \rangle$ that we obtained for $\mathcal{I}$. This reasoning shows that we can use the induction hypothesis for $\hat{\mathcal{I}}$: if $\boldsymbol{u}, \boldsymbol{v} \in \mathrm{dom}\,g$ and $(\boldsymbol{u}, \boldsymbol{v}) \prec (\boldsymbol{x}, \boldsymbol{y})$ then $g(\boldsymbol{u} \sqcap \boldsymbol{v}) + g(\boldsymbol{u} \sqcup \boldsymbol{v}) \leq g(\boldsymbol{u}) + g(\boldsymbol{v})$.

Let $\hat{\boldsymbol{x}}, \hat{\boldsymbol{y}}, \hat{\boldsymbol{x}}', \hat{\boldsymbol{y}}'$ be restrictions of respectively $\boldsymbol{x}, \boldsymbol{y}, \boldsymbol{x}', \boldsymbol{y}'$ to $\hat{V}$. We can write

$$\begin{aligned} g(\hat{\boldsymbol{y}}) = g(\hat{\boldsymbol{y}}') &= u(y_s) + f(y_s, \hat{\boldsymbol{y}}) = f(\boldsymbol{y}) + C \qquad \text{(since } (x_s, \hat{\boldsymbol{y}}) = \boldsymbol{y}' \notin \mathrm{dom}\,f\text{)} \\ g(\hat{\boldsymbol{x}}) &= f(x_s, \hat{\boldsymbol{x}}) = f(\boldsymbol{x}) \end{aligned}$$

By the induction hypothesis

$$g(\hat{\boldsymbol{x}} \sqcap \hat{\boldsymbol{y}}) + g(\hat{\boldsymbol{x}} \sqcup \hat{\boldsymbol{y}}) \leq g(\hat{\boldsymbol{x}}) + g(\hat{\boldsymbol{y}}) = f(\boldsymbol{x}) + f(\boldsymbol{y}) + C \tag{40}$$

We have $g(\hat{\boldsymbol{x}} \sqcup \hat{\boldsymbol{y}}) < \infty$, so we must have either $g(\hat{\boldsymbol{x}} \sqcup \hat{\boldsymbol{y}}) = f(x_s, \hat{\boldsymbol{x}} \sqcup \hat{\boldsymbol{y}})$ or $g(\hat{\boldsymbol{x}} \sqcup \hat{\boldsymbol{y}}) = f(y_s, \hat{\boldsymbol{x}} \sqcup \hat{\boldsymbol{y}}) + C = f(\boldsymbol{x} \sqcup \boldsymbol{y}) + C$. The former case is impossible since $(x_s, \hat{\boldsymbol{x}} \sqcup \hat{\boldsymbol{y}}) = \boldsymbol{x}' \notin \mathrm{dom}\,f$, so $g(\hat{\boldsymbol{x}} \sqcup \hat{\boldsymbol{y}}) = f(\boldsymbol{x} \sqcup \boldsymbol{y}) + C$. Combining it with (40) gives

$$g(\hat{\boldsymbol{x}} \sqcap \hat{\boldsymbol{y}}) + f(\boldsymbol{x} \sqcup \boldsymbol{y}) \leq f(\boldsymbol{x}) + f(\boldsymbol{y}) \tag{41}$$

This implies that $g(\hat{\boldsymbol{x}} \sqcap \hat{\boldsymbol{y}}) < C$, so we must have $g(\hat{\boldsymbol{x}} \sqcap \hat{\boldsymbol{y}}) = f(x_s, \hat{\boldsymbol{x}} \sqcap \hat{\boldsymbol{y}}) = f(\boldsymbol{x} \sqcap \boldsymbol{y})$. Thus, (41) is equivalent to (36).

**Case 2** $|Y| \geq 2$ or $|Y| = 1, \delta(\boldsymbol{x}, \boldsymbol{x} \sqcup \boldsymbol{y}) = 0$. It can be checked that $\boldsymbol{x} \sqcap (\boldsymbol{x} \sqcup \boldsymbol{y}) = \boldsymbol{x}$. Therefore, if we define $\boldsymbol{x}' = \boldsymbol{x}, \boldsymbol{y}' = \boldsymbol{x} \sqcup \boldsymbol{y}$ then the following identities hold:

$$\boldsymbol{x}' \sqcap \boldsymbol{y} = \boldsymbol{x} \sqcap \boldsymbol{y} \qquad \boldsymbol{x}' \sqcup \boldsymbol{y} = \boldsymbol{y}' \qquad \boldsymbol{x} \sqcap \boldsymbol{y}' = \boldsymbol{x}' \qquad \boldsymbol{x} \sqcup \boldsymbol{y}' = \boldsymbol{x} \sqcup \boldsymbol{y} \tag{42}$$

Let us select node $s \in Y$ and modify $\boldsymbol{x}'$ by setting $x'_s = y_s$. (Note that we have $y'_s = y_s$.) It can be checked that (42) still holds. We have $(\boldsymbol{x}', \boldsymbol{y}) \prec (\boldsymbol{x}, \boldsymbol{y})$ and $(\boldsymbol{x}, \boldsymbol{y}') \prec (\boldsymbol{x}, \boldsymbol{y})$ since $\Delta(\boldsymbol{x}', \boldsymbol{y}) = \Delta(\boldsymbol{x}, \boldsymbol{y}) - \{s\}$ and $\Delta(\boldsymbol{x}, \boldsymbol{y}') = \Delta(\boldsymbol{x}, \boldsymbol{y}) - (Y - \{s\})$, so by the induction hypothesis

$$f(\boldsymbol{x} \sqcap \boldsymbol{y}) + f(\boldsymbol{y}') \leq f(\boldsymbol{x}') + f(\boldsymbol{y}) \tag{43}$$

assuming that $\boldsymbol{x}' \in \mathrm{dom}\,f$, and

$$f(\boldsymbol{x}') + f(\boldsymbol{x} \sqcup \boldsymbol{y}) \leq f(\boldsymbol{x}) + f(\boldsymbol{y}') \tag{44}$$

assuming that $y' \in \text{dom} f$. If $x' \in \text{dom} f$ then Inequality (43) implies that $y' \in \text{dom} f$, and the claim then follows from summing (43) and (44). We now assume that $x' \notin \text{dom} f$; Inequality (44) then implies that $y' \notin \text{dom} f$.

Assume for simplicity of notation that $k$ corresponds to the first argument of $f$. Define instance $\hat{\mathcal{I}}$ with the set of nodes $\hat{V} = V - \{s\}$ and cost function

$$g(\boldsymbol{z}) = \min_{a \in D_s} \{u(a) + f(a, \boldsymbol{z})\} \qquad \forall \boldsymbol{z} \in \hat{\mathcal{D}} \equiv \bigotimes_{i \in \hat{V}} D_i$$

where $u(a)$ is the following unary term: $u(y_s) = 0$, $u(x_s) = C$ and $u(a) = \infty$ for $a \in D - \{x_s, y_s\}$. Here $C$ is a sufficiently large constant, namely $C > f(\boldsymbol{x}) + f(\boldsymbol{y})$.

Let $\hat{\boldsymbol{x}}, \hat{\boldsymbol{y}}, \hat{\boldsymbol{x}}', \hat{\boldsymbol{y}}'$ be restrictions of respectively $\boldsymbol{x}, \boldsymbol{y}, \boldsymbol{x}', \boldsymbol{y}'$ to $\hat{V}$. We can write

$$\begin{aligned} g(\hat{\boldsymbol{x}}) = g(\hat{\boldsymbol{x}}') &= u(x_s) + f(x_s, \hat{\boldsymbol{x}}) = f(\boldsymbol{x}) + C \qquad \text{(since } (y_s, \hat{\boldsymbol{x}}) = \boldsymbol{x}' \notin \text{dom} f) \\ g(\hat{\boldsymbol{y}}) &= f(y_s, \hat{\boldsymbol{y}}) = f(\boldsymbol{y}) \end{aligned}$$

By the induction hypothesis

$$g(\hat{\boldsymbol{x}} \sqcap \hat{\boldsymbol{y}}) + g(\hat{\boldsymbol{x}} \sqcup \hat{\boldsymbol{y}}) \leq g(\hat{\boldsymbol{x}}) + g(\hat{\boldsymbol{y}}) = f(\boldsymbol{x}) + f(\boldsymbol{y}) + C \tag{45}$$

We have $g(\hat{\boldsymbol{x}} \sqcup \hat{\boldsymbol{y}}) < \infty$, so we must have either $g(\hat{\boldsymbol{x}} \sqcup \hat{\boldsymbol{y}}) = f(y_s, \hat{\boldsymbol{x}} \sqcup \hat{\boldsymbol{y}})$ or $g(\hat{\boldsymbol{x}} \sqcup \hat{\boldsymbol{y}}) = f(x_s, \hat{\boldsymbol{x}} \sqcup \hat{\boldsymbol{y}}) + C = f(\boldsymbol{x} \sqcup \boldsymbol{y}) + C$. The former case is impossible since $(y_s, \hat{\boldsymbol{x}} \sqcup \hat{\boldsymbol{y}}) = \boldsymbol{y}' \notin \text{dom} f$, so $g(\hat{\boldsymbol{x}} \sqcup \hat{\boldsymbol{y}}) = f(\boldsymbol{x} \sqcup \boldsymbol{y}) + C$. Combining it with (45) gives

$$g(\hat{\boldsymbol{x}} \sqcap \hat{\boldsymbol{y}}) + f(\boldsymbol{x} \sqcup \boldsymbol{y}) \leq f(\boldsymbol{x}) + f(\boldsymbol{y}) \tag{46}$$

This implies that $g(\hat{\boldsymbol{x}} \sqcap \hat{\boldsymbol{y}}) < C$, so we must have $g(\hat{\boldsymbol{x}} \sqcap \hat{\boldsymbol{y}}) = f(y_s, \hat{\boldsymbol{x}} \sqcap \hat{\boldsymbol{y}}) = f(\boldsymbol{x} \sqcap \boldsymbol{y})$. Thus, (46) is equivalent to (36).